\documentclass[11pt]{article}
\usepackage{fullpage}
\usepackage{graphicx}
\usepackage{latexsym,amsmath,amssymb,amsthm,epic,eepic,multirow}
\usepackage{natbib}

\usepackage{multirow}
\usepackage{subfigure} 
\usepackage{natbib}
\usepackage{algorithm}
\usepackage{algorithmic}
\usepackage{mdframed}
\usepackage{tikz}
\usepackage{hyperref}
\usepackage{listings} 
\usepackage{color} 
\usepackage{texab}
\usepackage{bbm}
\usepackage{relsize}
\usepackage{array}
\newcolumntype{P}[1]{>{\centering\arraybackslash}p{#1}}

\definecolor{upmaroon}{rgb}{0.48, 0.07, 0.07}
\definecolor{royalazure}{rgb}{0.0, 0.22, 0.66}
\definecolor{pakistangreen}{rgb}{0.0, 0.4, 0.0}

\newcommand{\EE}{\mathbb{E}}

\theoremstyle{definition}
\newtheorem{theo}{Theorem}

\newtheorem{lemm}{Lemma}

\usepackage{color}

\let\originalleft\left
\let\originalright\right
\renewcommand{\left}{\mathopen{}\mathclose\bgroup\originalleft}
\renewcommand{\right}{\aftergroup\egroup\originalright}
\makeatletter
\newcommand{\leqnomode}{\tagsleft@true}
\newcommand{\reqnomode}{\tagsleft@false}
\makeatother

\begin{document}

\title{Multicarving for high-dimensional post-selection inference}  
  
\author{Christoph Schultheiss, Claude Renaux and Peter B\"uhlmann\\
Seminar for Statistics, ETH Z\"urich}

\maketitle

\begin{abstract}
We consider post-selection inference for high-dimensional (generalized)
linear models. Data carving \citep{fithian2014optimal} is a promising
technique to perform this task. However, it suffers from the instability of the
model selector and hence, may lead to poor replicability, especially in 
high-dimensional settings. We propose the multicarve method inspired by
multisplitting to improve upon stability and replicability. Furthermore,
we extend existing concepts to group inference and illustrate the
applicability of the methodology also for generalized linear models. 
\end{abstract}

\section{Introduction} \label{intro}
We consider post-selection inference in high-dimensional (generalized)
linear models. Statistical inference in high-dimensional models is
challenging: in a frequentist setting, the main methods use some
bias-corrected estimators of the Lasso \citep{zhang2014confidence, van2014asymptotically, javanmard2014confidence} or of Ridge regression
\citep{buhlmann2013statistical}, and \cite{cai2017confidence}
provide refined optimality results for such techniques. On the other hand,
post-selection inference provides a very different approach for
constructing confidence statements in high-dimensional
models. Post-selection inference is attractive as it is closer in some
vague sense to what practitioners like to do, namely to apply first some model
selection in order to restrict the set of covariates and make the problem
feasible. Post-selection inference has long been viewed as rather ill-posed \citep{leeb2003finite} until \cite{berk2013valid} provided a conservative
approach to improve its image. More recent work by \cite{fithian2014optimal}, \cite{tian2018selective}, \cite{taylor2018post} and others lead to interesting new inferential tools. The
current work is building on those contributions. 

\paragraph{The instability of post-selection inference.} Post-selection
inference deals with the problem of inference statements, after having
selected a set of covariates using a data-driven algorithm or method. For
post-selection 
inference in high-dimensional (generalized) linear 
models, a very popular model selection method is the Lasso
\citep{tibshirani1996regression}; and in fact, in this work, we only focus
on the Lasso as model selector. Among the main concerns when using the
Lasso or any other variable selection method is its \emph{instability}. The selected model, say, by the Lasso, has low
degree of replicability due to its instability arising from correlated
covariates and/or high noise scenarios. Thus, the inference after model
selection might be very non-replicable if the model selector leads to
different results for small perturbations of the data. Take getting
new realizations from the same data generating process as an example. Our new
multicarving proposal is a possible remedy to make post-selection inference
more reproducible. 

\medskip 
A variety of approaches to get valid tests and confidence intervals after model selection have been developed. In order to put our proposal in some context, we discuss briefly the ones most relevant to our work in the following.

A simple approach for valid inference is to split the data into two parts and use the first half for selection and the second half for inference \citep{wasserman2009high}. Thus, the idea is very similar to any validation scheme using data splitting. 

This simple single data splitting method has certain drawbacks. Since
splitting the data is a 
random process, the inference statements change if a different split is
chosen. If we repeat this process multiple times, we observe that the
obtained p-values per predictor change a lot: \cite{meinshausen2009p} call this
phenomenon the ``p-value lottery''. For the Lasso selector, this is
especially accentuated as it is highly non-stable and potentially selects
quite different models depending on the split. Therefore, results obtained
through this method are not replicable at all unless one fixes the
split. In order to receive more stable and replicable p-values,
\cite{meinshausen2009p} suggest splitting the data multiple times, say,
$B=50$ times leading to 
p-values $P_j^{\left(b\right)}$ for each split $b=1,\ldots,B$ and each
predictor $j=1,\ldots,p$. The p-values per predictor are aggregated using
quantile functions and adequate correction terms. Although there is still
randomness involved, the results should become more stable with increasing
$B$ in the spirit of the law of large numbers. This technique is referred
to as multisplitting. 

To avoid confusion, we save the term post-selection inference for techniques that perform inference on the same data as used for selection and refer to the methods from \citet{wasserman2009high} and \citet{meinshausen2009p} as (multi)splitting. Post-selection inference for a (generalized) linear model can be achieved by calculating or simulating a constrained null distribution, where the constraints reflect the selected model.

\cite{lee2016exact} analyze the case of Lasso selection in a linear
model. They show that the Karush-Kuhn-Tucker (KKT) criteria, which are
necessary conditions for the Lasso solution, lead to a polyhedral
constraint on the observed response vector. Using this constraint, they
derive a truncated normal distribution which allows for valid inference. A
drawback of this method is a loss in power introduced by those polyhedral
constraints. Similar constraints have been derived in
\cite{tibshirani2016exact} for sequential regression problems: compared
to Lasso selection for fixed value of $\lambda$, those constraints
increase in dimensionality rather quickly, since every step of the
procedure results in additional constraints.

Somewhere in between data splitting and post-selection inference is a
technique called data carving \citep{fithian2014optimal}. In order to
distinguish data carving from methods as in \cite{lee2016exact}, we refer to the latter as pure post-selection inference in the following. Due
to the loss in power introduced by pure post-selection inference,
\cite{fithian2014optimal} prefer not to use all observations for the
selection process. Further, they prove that completely discarding the
fraction of data used for model selection in the inference stage leads to
inadmissible tests. Instead, one should use as much information of the
selection data as is still usable and should only discard the information that
was actually needed to obtain the given selection. This means that one
``carves'' the data. One can reuse the selection constraints introduced for
pure post-selection inference but imposes them on the selection data
only. This method outperforms pure post-selection inference and simple
sample splitting with respect to power. Though, it is computationally much
more involved under certain model assumptions. Naturally, pure
post-selection inference can be seen as a special case of data carving, and
\cite{fithian2014optimal} refer to it as Carve$_{100}$. 

\cite{barber2015controlling} introduce the concept of knockoff filters for model selection and inference. Their main idea is to compare the measurable effect of the regressor covariates to the corresponding effect of their ``knockoff copies'' which should behave statistically equivalent for covariates with no true underlying effect. \cite{barber2019knockoff} adapt this methodology to the high-dimensional setting and post-selection inference. The data is split into two parts for that purpose, one for selection and one for inference only. The authors also suggest a method which can ``recycle'' some of the information from the selection data in the inference stage, which resembles the data carving idea. However, they condition not only on the selection event but on the full observation of the selection data. This has the advantage that the selection process on the first part of the data can be arbitrarily and is not restricted to methods for which one can sample from the data conditional on the selection event.

\cite{berk2013valid} provide an inference technique that is valid given any preceding model selection procedure, potentially, inspecting all of the data. This is possible by using the so-called PoSI (post-selection inference) constant $K$. This constant is defined as the minimal value such that the maximal absolute t-statistic maximized overall possible predictor variables and submodels is at most equal to $K$ with probability at least $1-\alpha$. The advantage of this method is that it leaves all freedom to the practitioner for the selection process without losing validity. For example, visual inspection of the data through a human, which is done quite often in practice, is allowed. On the other hand, this method is quite conservative by construction. Furthermore, calculating the constant $K$ gets computationally involved such that the authors only suggest to use their method for up to $p \approx 20$. Despite the nice theoretical framework, the method is not suited for high-dimensional statistics, which is our focus.

Recent developments by \cite{kuchibhotla2020valid} lead to computationally efficient procedures with similar guarantees. They derive a method to construct confidence regions such that they contain the true parameter in any submodel simultaneously with probability at least $1-\alpha$. Due to this simultaneous coverage any possible model selection can be applied and the true parameter is still contained in the constructed region. Naturally, this method is also rather conservative. Especially, it cannot gain power from a sparsity assumption due to the simultaneous coverage in all submodels.

\subsection{Relation to other work and contribution}
\cite{meinshausen2009p} as well as \cite{fithian2014optimal} emphasize different drawbacks of the simple idea of data-splitting for inference in high-dimensional statistics and show how to improve on them. Therefore, we focus on how to optimally combine those improvements leading to our ``multicarving'' method. Since we work with the Lasso as model selector, we also build on the results of \cite{lee2016exact}.

We further elaborate two more extensions of data carving in a linear model
that can be combined with multisplitting in the same fashion. The first one
concerns group testing. There are many developments in high-dimensional
statistics for testing groups of covariates for significance instead of
single covariates, see for example \cite{van2016chi}, \cite{mitra2016benefit}, 
and \cite{guo2019group}. Group tests are of 
particular use as with many (highly correlated) covariates, it might be
overly ambitious to correctly detect the individual active variables,
whereas groups of variables might be more realistic to detect. Hierarchical
testing schemes are particularly attractive for this task; see for example
\cite{manpb16} and \cite{renaux2018hierarchical}. Secondly, we provide
extensions of multicarving to generalized linear models. Pure post-selection
inference in logistic linear regression is discussed in
\cite{taylor2018post} who 
rely on asymptotic Gaussianity. As for the linear model, pure
post-selection inference is suboptimal regarding power, thus, we extend
their argument to the data carving approach. We only provide a detailed
discussion for the case of logistic linear regression. Though, similar
adjustments could be done for other generalized linear models.

\section{Methodology for high-dimensional post-selection inference}\label{method}
We first consider the methodological framework for linear models and summarize multisplitting (Section \ref{ms}) as well as data carving (Section \ref{datacarving}). This serves as a basis to develop our novel multicarving procedure for single covariates (Section \ref{mdc}) and an extension to group inference (Section \ref{group}) and logistic regression or other generalized linear models (Section \ref{GLM}). While those developments focus on hypothesis testing, we discuss confidence intervals in Section \ref{CI}.
\subsection{High-dimensional linear model and inference for single variables} \label{linmod}
We assume to have a response vector $\mathbf{Y}=\left(Y_1,\ldots,Y_n\right)^{\top}$ and a (fixed) design matrix $X \in \mathbb{R}^{n\times p}$, where $p \gg n$. This yields a linear model of the form 
\begin{equation} \label{eq:linmod}
\mathbf{Y}=X\boldsymbol{\beta}+\boldsymbol{\epsilon},
\end{equation}
where $\boldsymbol{\epsilon}=\left(\epsilon_1,\ldots,\epsilon_n\right)^{\top}$ consists of i.i.d. $\mathcal{N}\left(0,\sigma^2\right)$ entries with known or unknown variance $\sigma^2$ and $\boldsymbol{\beta} \in \mathbb{R}^p$ is the unknown parameter of interest. We represent vectors in boldface, whereas scalars and matrices are written in usual letters. We write $\mathbf{y}$ for a given realization of the random vector $\mathbf{Y}$. We use index $1$ ($X_1$,$\mathbf{Y}_1$ and $\mathbf{y}_1$) and index $2$ ($X_2$,$\mathbf{Y}_2$ and $\mathbf{y}_2$) to denote selection data and data used for inference only, respectively. Further, we assume that the active set $S=\left\{j;\beta_j \neq 0\right\}$ is sparse, i.e.,\ $s=\left\vert S\right\vert \ll n$ such that inference using ordinary least squares would be possible on the data if the true active set was known.

After data-driven model selection, we deal with a subset $\tilde{S}$ of size $\tilde{s}=\big\vert\tilde{S}\big\vert$. We aim to perform inference based on this subset $\tilde{S}$. We write $X_{\tilde{S}}$ for the matrix $X$ restricted to the selected columns. Likewise, $X_{1,\tilde{S}}$ and $X_{2,\tilde{S}}$ denote selection and inference data restricted to the selected columns. Generally, a distinction has to be made whether we test 
\begin{equation} \label{eq:fullnull}
H_{0,j}: \quad \beta_j=0 \qquad \text{versus} \qquad H_{A,j}: \quad \beta_j \neq 0
\end{equation}
for the entries of the full $\boldsymbol{\beta} \in \mathbb{R}^p$ or if the test is made with respect to 
\begin{equation} \label{eq:selectednull}
H_{0,j}^{\tilde{S}}: \quad \beta_j^{\tilde{S}}=0 \qquad \text{versus} \qquad H_{A,j}^{\tilde{S}}: \quad \beta_j^{\tilde{S}} \neq 0.
\end{equation}
Here, $\boldsymbol{\beta}^{\tilde{S}} \in \mathbb{R}^{\tilde{s}}$ corresponds to the selected submodel and is defined as
\begin{equation} \label{eq:bestlin}
\boldsymbol{\beta}^{\tilde{S}} \ \equiv \ \underset{\mathbf{b}^{\tilde{S}}}{\text{arg min }}\EE\left\Vert\mathbf{Y}-X_{\tilde{S}}\mathbf{b}^{\tilde{S}}\right\Vert^2\ = \ X_{\tilde{S}}^+X\boldsymbol{\beta},
\end{equation}
the best linear predictor in the given model. We write $X_{\tilde{S}}^+$ for $\big(X_{\tilde{S}}^{\top} X_{\tilde{S}}\big)^{-1}X_{\tilde{S}}^\top$, i.e.,\ the generalized inverse of $X_{\tilde{S}}$. We introduce corresponding null hypotheses for groups of variables in Section \ref{group}.

Typically, an inference statement for \eqref{eq:fullnull} would be more favorable, since we are interested in the true underlying model. Though, tests for \eqref{eq:selectednull} are valid under weaker assumptions.

Of particular interest is the screening property. Screening is defined as $\tilde{S} \supseteq S$ or in words, screening asks for all active variables being part of the selected model. If this holds, we have $\beta_j^{\tilde{S}}=\beta_j \quad \forall j \in \tilde{S}$. Thus, tests valid for \eqref{eq:selectednull} are also unbiased for \eqref{eq:fullnull} assuming screening. Importantly, screening is a requirement on the initial model selection process and not on the following inference calculation.

We focus on model selection using the Lasso. The screening property for the Lasso is rather delicate to achieve in the finite sample case. Though, it can be guaranteed with probability $1$ for $n \rightarrow \infty$ under adequate conditions. Such conditions are discussed in \cite{meinshausen2006high}, \cite{meinshausen2009lasso} and \cite{bickel2009simultaneous}, see also the book by \cite{buhlmann2011statistics}.
\subsection{Previously proposed methods}
We first review some earlier work which
serves as a basis for our new proposal in
Section \ref{mdc}.
\subsubsection{Multisplitting for inference} \label{ms}
In this section, we briefly summarize the multisplitting method introduced in \cite{meinshausen2009p}. Multisplitting works as follows:\\
For each $b=1,\ldots,B$:
\begin{enumerate}
\item Randomly split the data into two disjoint groups of sizes $n_1$ and $n_2$.
\item Find $\tilde{S}^{\left(b\right)}$ using $X_1$ and $\mathbf{y}_1$.
\item For $j \in \tilde{S}^{\left(b\right)}$, calculate p-values $p_j^{\left(b\right)}$ using $X_{2,\tilde{S}^{\left(b\right)}}$ and $\mathbf{y}_2$ with ordinary least-squares; for $j \notin \tilde{S}^{\left(b\right)}$, set $p_j^{\left(b\right)}=1$.
\item Adjust the p-values to $P_j^{\left(b\right)}=\text{min}\left(p_j^{\left(b\right)}\tilde{s}^{\left(b\right)},1\right)$ to correct for multiplicity using Bonferroni adjustment.
\end{enumerate}
The fourth step is designed to control the family-wise error rate (FWER). Throughout this work, we use lower case letters ($p$) for raw p-values that result from a test and upper case letters ($P,Q$) for p-values resulting from any correction or aggregation. The default value for splitting is $n_1 = \left \lfloor{\tfrac{n}{2}}\right \rfloor$. It remains to aggregate the $B$ p-values for covariate $j$. Valid aggregation is possible by using a quantile of fixed fraction $\gamma \in \left(0,1\right]$ as
\begin{equation} \label{eq:fixquant}
Q_j\left(\gamma\right)=\text{min}\left\{1,\ q_{\gamma}\left(\left\{P_j^{\left(b\right)}/\gamma; \ b=1, \ldots ,B\right\}\right)\right\},
\end{equation}
with $q_{\gamma}$ being the empirical quantile function. Since a good choice of $\gamma$ might not be known a priori, one can also optimize $\gamma$ over a range $\left[\gamma_\text{min},1\right]$ where $\gamma_\text{min}\in \left(0,1\right]$. This yields a different p-value
\begin{equation} \label{eq:optquant}
P_j=\text{min}\Big\{1,\ \left(1-\text{log}\left(\gamma_\text{min}\right)\right)\underset{\gamma \in \left[\gamma_\text{min},1\right]} {\text{min}} Q_j\left(\gamma\right)\Big\}.
\end{equation}
The additional factor $\left(1-\text{log}\left(\gamma_\text{min}\right)\right)$ corrects for optimizing over all possible quantiles. A typical choice is $\gamma_\text{min}=0.05$, yielding a correction factor of $\left(1-\text{log}\left(0.05\right)\right)\approx 3.996$.

Without any screening assumption, those p-values actually test the following null hypothesis for some given covariate $j$
\begin{equation} \label{eq:multiplenull}
H_{0,j}^{\tilde{S}^{\left(1\right)},\ldots,\tilde{S}^{\left(B\right)}}: \quad \beta_j^{\tilde{S}^{\left(b\right)}}=0 \quad \forall b \qquad \text{versus} \qquad H_{A,j}^{\tilde{S}^{\left(1\right)},\ldots,\tilde{S}^{\left(B\right)}}: \quad \exists b \text{ s.t.} \ \beta_j^{\tilde{S}^{\left(b\right)}} \neq 0.
\end{equation}

Given two conditions, \cite{meinshausen2009p} derive asymptotic (for $n\rightarrow \infty$) FWER control with respect to null hypothesis \eqref{eq:fullnull}. The conditions are:
\leqnomode
\begin{flalign}
& \qquad \text{ Asymptotic screening: lim}_{n\rightarrow \infty}\PR{\tilde{S}\supseteq S}=1. && \tag{A1} \label{ass:1}\\
& \qquad \text{ Sparsity: }\tilde{s}< n_2 .&& \tag{A2}\label{ass:2}
\end{flalign}
\reqnomode
The screening condition, as argued before, leads to $\beta_j^{\tilde{S}^{\left(b\right)}}=\beta_j \quad \forall j \in \tilde{S}^{\left(b\right)}$ and makes the inference statement valid for the true underlying parameter vector. The sparsity condition enables us to do least-squares inference, implicitly assuming that $X_{2,\tilde{S}^{\left(b\right)}}$ has full column rank for all $b$.

If screening held in the finite sample case as well, the error control could be formulated in a non-asymptotic sense. Although this is usually not the case, the simulations in \cite{meinshausen2009p} as well as ours show that multisplitting controls the type-I error with respect to \eqref{eq:fullnull} clearly better than single-splitting when screening cannot be guaranteed. This can be explained by the ``p-value lottery'': 
Every split results in different p-values for the selected variables. 
There are chances that some true non-active variables are significant for some splits. After aggregation, only variables that are significant in a decent number of splits remain significant overall. Due to the variability of these p-values over different splits, chances are that fewer non-active variables get rejected after aggregation than in the average single split. Thus, multisplitting leads to better error control.

\subsubsection{Data carving} \label{datacarving}
In this section, we discuss the idea of data carving introduced in \cite{fithian2014optimal}. We focus on the special case of the linear model \eqref{eq:linmod} with Lasso selection, which we will later extend to logistic regression and other generalized linear models. We emphasize that they provide a theoretical framework that could be applied to a much broader spectrum of problems.

The main conceptual idea of data carving is summarized in the following statement \citep{fithian2014optimal}: ``The answer must be valid, given that the question was asked.'' Thus, one should control the selective type-I error rate
\begin{equation} \label{eq:selerror}
\mathrm{P}_{H_0^{\tilde{S}}}\left[\text{reject} \quad H_0^{\tilde{S}}\ \Big| \left(\tilde{S},H_0^{\tilde{S}}\right) \ \text{selected}\right]\leq \alpha.
\end{equation}
The hypothesis $H_0^{\tilde{S}}$ is a general notation for a hypothesis as, e.g.,\ in \eqref{eq:selectednull}. Define the event $M\left(\mathbf{Y}_1\right)$ as $\big\{\big(\tilde{S},H_0^{\tilde{S}}\big) \ \text{selected}\big\}$, the selection event using data $\left\{X_1,\mathbf{Y}_1\right\}$. We write $M\left(\mathbf{Y}_1\right)$ since $X_1$ is assumed to be fixed. Then, the requirement \eqref{eq:selerror} can be equivalently stated as
\begin{equation} \label{eq:selectiveerror}
\mathrm{P}_{H_0^{\tilde{S}}}\left[\text{reject} \ H_0^{\tilde{S}}\ \Big| M\left(\mathbf{Y}_1\right) \right]\leq \alpha.
\end{equation} 
Simple data splitting on the other hand controls the following error
\begin{equation*}
\mathrm{P}_{H_0^{\tilde{S}}}\left[\text{reject} \ H_0^{\tilde{S}}\ \Big| \mathbf{Y}_1 \right]
\end{equation*}
at level $\alpha$. Thus, more conditioning is done than would theoretically be needed, since $M\left(\mathbf{Y}_1\right)$ does not contain all information about $\mathbf{Y}_1$ but only guarantees that it results in the observed selection event.

To perform inference controlling the error in \eqref{eq:selectiveerror}, one needs to understand the distribution of $\mathbf{Y}\ \big| M\left(\mathbf{Y}_1\right)$. The first step is to understand the selection event $ M\left(\mathbf{Y}_1\right)$. We focus on our case of interest, inference in the linear model \eqref{eq:linmod} using Lasso selection. More precisely, let Lasso selection be defined as follows
\begin{equation} \label{eq:lasso}
\widehat{\boldsymbol{\beta}}=\underset{\boldsymbol{\beta}}{\text{arg min}}\ \tfrac{1}{2}\left\Vert\mathbf{y}_1-X_1\boldsymbol{\beta}\right\Vert_2^2+\lambda\left\Vert\boldsymbol{\beta}\right\Vert_1\\
\end{equation}
\begin{equation*}
\tilde{S}=\left\{j:\widehat{\beta}_j\neq0\right\}.
\end{equation*}
There exist different definitions of the Lasso that are equivalent after rescaling. We use definition \eqref{eq:lasso} following \cite{lee2016exact} where this selection event is fully characterized. The set of $\mathbf{Y}_1$ that would lead to the same $\tilde{S}$ forms a union of polyhedra in $\mathbb{R}^{n_1}$. If we additionally condition on the signs of the parameters' Lasso estimates, $\text{sign}\big(\widehat{\beta}_j\big)\ \forall j \in \tilde{S}$, this union is shrunk to a single polyhedron. Dealing with a single polyhedron is easier both computationally as well as from a theoretical perspective. Hereafter, we additionally condition on the signs at the price of a small loss in power. This single polyhedron can easily be described by linear inequality constraints, e.g.,\ $A\mathbf{Y}\leq\mathbf{b}$. Those constraints can be split into ``active'' ($A_1\mathbf{Y}\leq\mathbf{b}_1$) and ``inactive'' ($A_0\mathbf{Y}\leq\mathbf{b}_0$) constraints which define statistically independent events. Further, $X_{\tilde{S}}^+\mathbf{Y}$ is independent of the inactive constraints such that it is also independent while conditioning on the active constraints, i.e.,\ $\left(X_{\tilde{S}}^+\mathbf{Y}\ \big| A_1\mathbf{Y}\leq\mathbf{b}_1\right)\perp \left( A_0\mathbf{Y}\leq\mathbf{b}_0\right)$. Therefore, we can ignore the inactive constraints for inference purposes which are based on $X_{\tilde{S}}^+\mathbf{Y}$. For simplicity, we refer to $A\mathbf{Y}\leq\mathbf{b}$ as being the active constraints only.

\cite{fithian2014optimal} elaborate how to handle $\mathbf{Y}\ \big| M\left(\mathbf{Y}_1\right)=\mathbf{Y}\ \big|A\mathbf{Y}\leq\mathbf{b}$ in a given model. As $\boldsymbol{\beta}^{\tilde{S}}$ is unknown, the conditional distribution is not tractable yet. To deal with this problem, one can treat the unknown parameters as nuisance parameters in an exponential family which one can get rid of by conditioning accordingly. Generally, one has to decide between the ``saturated model'' and the ``selected model'':
\begin{itemize}
\item Saturated model: $\boldsymbol{\mu}=\EE\left[\mathbf{Y}\right]$ has $n$ degrees of freedom and $\boldsymbol{\beta}^{\tilde{S}}=X_{\tilde{S}}^+\boldsymbol{\mu}$ is the best linear predictor based on the selected model (cf.\ Equation \eqref{eq:bestlin}).
\item Selected model: $\boldsymbol{\mu}=\EE\left[\mathbf{Y}\right]=X_{\tilde{S}}\boldsymbol{\beta}^{\tilde{S}}$ has $\tilde{s}$ degrees of freedom and $\boldsymbol{\beta}^{\tilde{S}}$ completely defines the distribution.
\end{itemize}
If we consider the saturated model, which includes more parameters than the selected model, more conditioning has to be done. This leads to a drop in power but with the advantage that tests are valid for \eqref{eq:selectednull} without any screening assumption. The selected model view is generally more powerful since less conditioning is done but it needs stronger assumptions to hold. The existence of $\boldsymbol{\beta}^{\tilde{S}}$ such that $\EE\left[\mathbf{Y}\right]=X_{\tilde{S}}\boldsymbol{\beta}^{\tilde{S}}$ is exactly the screening condition. If screening holds, either approach is valid to test \eqref{eq:fullnull}. Since we are mainly interested in this null hypothesis, we focus on the selected model leading to more powerful tests under screening. In Section \ref{CI}, we elaborate further on the saturated method and its advantages.

Consider the selected model. To perform inference for covariate $j$, one has to condition onto $\big(X_{\tilde{S}\setminus j}\big)^{\top}\mathbf{Y}$. After applying this conditioning, the random vector of interest 
$\big(X_{\tilde{S}}^+\big)_j\mathbf{Y}$ is independent from the unknown parameters $\boldsymbol{\beta}^{\tilde{S}}_{-j}$. This leads to a degenerate truncated multivariate Gaussian distribution with no more unknown nuisance parameters. The truncation is defined by the selection event. To test the null hypothesis, one further assumes $\beta^{\tilde{S}}_j=0$. Thus, one is interested in 
\begin{equation} \label{eq:carvepv}
p_j\left(\mathbf{y}\right)=\begin{cases}
\PR{\left(X_{\tilde{S}}^+\right)_j\mathbf{Y}\geq \left(X_{\tilde{S}}^+\right)_j\mathbf{y}\ \Big| \beta^{\tilde{S}}_j=0, \left(X_{\tilde{S}\setminus j}\right)^{\top}\mathbf{Y}=\left(X_{\tilde{S}\setminus j}\right)^{\top}\mathbf{y},A\mathbf{Y}\leq\mathbf{b}} & \text{if } \widehat{\beta}_j>0\\[10pt]
\PR{\left(X_{\tilde{S}}^+\right)_j\mathbf{Y}\leq \left(X_{\tilde{S}}^+\right)_j\mathbf{y}\ \Big| \beta^{\tilde{S}}_j=0, \left(X_{\tilde{S}\setminus j}\right)^{\top}\mathbf{Y}=\left(X_{\tilde{S}\setminus j}\right)^{\top}\mathbf{y},A\mathbf{Y}\leq\mathbf{b}} & \text{if } \widehat{\beta}_j<0.
\end{cases}
\end{equation}
Note that we can use one-sided tests, since we implicitly condition on the sign of $\widehat{\beta}_j$ by restricting ourselves to the single polyhedron $A\mathbf{Y}\leq\mathbf{b}$. If we have selected a correct model such that the selected model view is applicable, $\sigma$ is known, and $j \notin S$ is not a true active variable, then we have $p_j\left(\mathbf{Y}\right) \sim \text{Unif}\left[0,1\right]$.
This null distribution is not easily tractable and thus the probability is hard to calculate. Though, it can be sampled from using MCMC. 
This means that data carving achieves higher power compared 
to sample splitting at the price of a substantially higher computational cost. We present an applicable MCMC sampling scheme in Appendix \ref{app:mcmc}.

In the saturated viewpoint, only one degree of freedom remains after conditioning (cf.\ Section \ref{CI}). Therefore, one can deal with a univariate truncated normal such that the exact probability, the analogue of \eqref{eq:carvepv}, can be calculated efficiently using the CDF of a Gaussian. Thus, the trade-off between the selected and the saturated model also involves a computational component.

\bigskip
So far, we assumed $\sigma$ to be known. If this is not the case, $\sigma^2$ could be handled as further nuisance parameter, which is resolved by additionally conditioning on $\left\Vert\mathbf{Y}\right\Vert^2$. However, this nonlinear constraint disables some of the computational shortcuts which all linear constraints allow for. In our simulations, we use some estimate $\widehat{\sigma}$ wherever the variance is assumed to be unknown and proceed as if it was known initially. For completeness, we mention that the distribution when additionally conditioning on $\left\Vert\mathbf{Y}\right\Vert^2$ is not Gaussian anymore. The corresponding null distribution can still be approximated using a different MCMC sampling technique. Note that this is only possible for the selected model. In the saturated model, one would end up imposing one quadratic and $n-1$ linear equality constraints onto an $n$-dimensional vector. This would only leave two points to sample from such that no inference is possible.

\subsection{Novel multicarving for valid inference} \label{mdc}
\cite{meinshausen2009p} have theoretically argued and empirically shown that splitting several times and aggregating is to be preferred over a single-split approach. On the other hand, \cite{fithian2014optimal} have shown that discarding all selection data in a splitting set-up is mathematically inadmissible and typically less efficient. To overcome this problem, they introduce the idea of data carving. Nevertheless, their approach potentially suffers from a similar p-value lottery as discussed in \cite{meinshausen2009p} since it is initiated by randomly splitting the data into two disjoint groups of given sizes; one for selection and inference, and the other one for inference only. 
Therefore, it is often difficult to replicate. 
Thus, we advocate the idea of applying data carving multiple times in order to a) overcome the p-value lottery and b) avoid the proven inadmissibility of any splitting procedure. We use the following procedure:\\
For $b=1, \ldots , B$:
\begin{enumerate}
\item Randomly split the data into two disjoint groups of sizes $n_1$ and $n_2$.
\item Find $\tilde{S}^{\left(b\right)}$ using $X_1$ and $\mathbf{y}_1$ with Lasso selection.
\item For $j \in \tilde{S}^{\left(b\right)}$, calculate p-values $p_j^{\left(b\right)}$ for the given split and selected model according to \eqref{eq:carvepv}, for $j \notin \tilde{S}^{\left(b\right)}$, set $p_j^{\left(b\right)}=1$.
\item Adjust the p-values to $P_j^{\left(b\right)}=\text{min}\left(p_j^{\left(b\right)}\tilde{s}^{\left(b\right)},1\right)$ to correct for multiplicity using Bonferroni adjustment.
\end{enumerate}
As in multisplitting, we include the fourth step in order to control the FWER. Different corrections could be applied to obtain some less restrictive error control such as the false discovery rate (FDR) as discussed in \cite{meinshausen2009p}. There is a trade-off involved in choosing $n_1$ and $n_2$. The higher we set $n_1$, the higher the probability of screening gets, which is required for valid tests. 
On the other hand, more power remains for 
the second stage, namely, the inference calculation, for higher values of $n_2$. We empirically analyze this trade-off in our simulations in Section \ref{res}. 
To get one p-value per predictor, we use the same aggregation techniques as presented in Section \ref{ms}, resulting in a single p-value $Q_j\left(\gamma\right)$ or $P_j$. In our simulations, we focus on optimizing over the quantiles as described in \eqref{eq:optquant} instead of using a fixed predefined quantile $\gamma$. To distinguish the different methods, we call this procedure multicarving and the method described in Section \ref{datacarving} single-carving.
\subsection{Saturated view and confidence intervals}\label{CI}
Naturally, one wants to perform inference without the screening assumption. As mentioned in Section \ref{datacarving}, we can use the saturated model from \cite{fithian2014optimal} for this purpose. In the saturated view, we do not assume the selected submodel to completely define the mean parameter $\boldsymbol{\mu}$ but only to approximate it as in \eqref{eq:bestlin}. In order to get rid of the unknown parameters and create a tractable distribution, we have to condition on to $\mathit{P}^{\perp}_{\boldsymbol{\eta}}\mathbf{Y}=\mathit{P}^{\perp}_{\boldsymbol{\eta}}\mathbf{y}$. Here, we define $\boldsymbol{\eta} \equiv \big(X_{\tilde{S}}^+\big)_j$, leading to $\boldsymbol{\eta}^{\top}\boldsymbol{\mu}=\beta^{\tilde{S}}_j$. As $\mathit{P}^{\perp}_{\boldsymbol{\eta}}$ has rank $n-1$, there remains only one degree of freedom after conditioning, namely, in the direction of $\boldsymbol{\eta}$. Therefore, one deals with a univariate truncated Gaussian where the truncation comes from invoking the selection event $A\mathbf{Y}\leq \mathbf{b}$. Inference statements can be calculated efficiently using the CDF of a univariate Gaussian. A detailed explanation of this procedure can be found in \cite{lee2016exact}.

This can be done regardless of the quality of the selected submodel. Therefore, the saturated viewpoint leads to valid tests for null hypotheses \eqref{eq:selectednull} (single-carving) and \eqref{eq:multiplenull} (multicarving) without any screening assumption. However, if screening fails, the best linear predictor in the submodel is generally non-sparse. This means that there is no $j\in \tilde{S} \ \text{s.t.} \ \beta_{j}^{\tilde{S}}= 0$ and there cannot be any false positives with respect to those null hypotheses. Therefore, such tests for null hypotheses without any screening assumption are not of particular interest.

Nevertheless, those tests can be used to determine confidence intervals. As for any test, confidence intervals for multicarving can be found by inverting it. \cite{dezeure2015high} give a detailed explanation of how to compute confidence intervals for multisplitting. We refrain from giving a full theoretical result for our derived method, but remark that their construction does not require the individual p-values to origin from a sample splitting procedure as long as they are valid. Therefore, this approach can be directly adopted to multicarving by calculating carving p-values but keeping the remaining scheme the same. We focus on the construction without multiplicity correction. For covariate $j$, this leads to a $\left(1-\alpha\right)$-confidence interval (CI) such that
\begin{equation}\label{eq:cicoverage}
\PR{\beta_{j}^{\tilde{S}^{\left(b\right)}}\in \text{CI}\ \forall b}\geq 1-\alpha,
\end{equation}
where $\beta_{j}^{\tilde{S}^{\left(b\right)}}$ are defined through \eqref{eq:bestlin}. This is of particular interest when $\beta_{j}^{\tilde{S}^{\left(b\right)}}$ differ for different splits $b$. 
Therefore, it appears natural to omit the screening assumption and to adopt the saturated model for our confidence intervals.
Further, the use of the saturated model leads to more efficient computation.

We focus on two-sided confidence intervals for two reasons. First, having both a lower and an upper bound might be more informative for a practitioner. Second, $\text{sign}\big(\beta_j^{\tilde{S}^{\left(b\right)}}\big)$ is not necessarily the same for all splits $b$ in which covariate $j$ is selected such that combining different splits to a one-sided confidence interval is not appropriate. Thus, the confidence intervals in this case are not the exact inversion of the hypothesis tests.

Notably, if one were to apply simultaneous tests for different null hypotheses in the selected model, this could be done by just calculating a single MCMC chain and relying on the idea of importance sampling afterwards. However, to get a precise enough statement for such simultaneous tests, more MCMC samples might be required than for just calculating a p-value such that this extra statement is not for free.

\subsection{Extension to group testing} \label{group}
In a high-dimensional set-up with potentially correlated predictors, finding individual active variables is often too ambitious. Especially, the Lasso selector struggles with distinguishing between two or more highly correlated variables. Therefore, one might prefer to test several variables as a group. We define the null hypothesis for a given group $G$ as
\begin{equation} \label{eq:fullnullgroup}
H_{0,G}: \quad \beta_j=0 \ \forall j\in G \qquad \text{versus} \qquad H_{A,G}: \quad \exists j\in G, \text{ s.t.} \ \beta_j \neq 0
\end{equation}
for the full model coefficients. Let $\tilde{G}=\tilde{S}\cap G$ be the variables in our group that have been selected then we define the null hypothesis in the selected model as
\begin{equation} \label{eq:selectednullgroup}
H_{0,G}^{\tilde{S}}: \quad \beta_j^{\tilde{S}}=0\ \forall j\in \tilde{G} \qquad \text{versus} \qquad H_{A,G}^{\tilde{S}}: \quad \exists j\in \tilde{G}, \text{ s.t.} \ \beta_j^{\tilde{S}} \neq 0.
\end{equation}

The practitioner often wants to test multiple groups or
test groups in a hierarchical fashion, say, in a
data-driven way. Of course, a multiplicity correction
has to be applied which is possible for
any valid group test which controls
the type I error. We refer to \cite{meinshausen2008hierarchical} for a
detailed explanation of a hierarchical testing procedure and corresponding
multiple-testing correction. 

\subsubsection{(Multi)splitting for group inference}
Groups of variables can be tested for significance in
the same way as single variables by splitting the data. The extension to groups follows naturally
as in the low-dimensional case by applying partial F-tests instead of
t-tests. This can be done either with a single split or multiple splits
using the previously mentioned aggregation techniques \eqref{eq:fixquant}
and \eqref{eq:optquant}.

\subsubsection{(Multi)carving for group inference}
The above mentioned (multi)splitting techniques for group inference suffer
from the same inadmissibility issue as in the single variable case as more
conditioning than necessary is applied. Therefore, we suggest a slight
transformation of the data carving idea which makes it applicable to
testing for group significance. We focus on the selected viewpoint meaning
that our derivation will actually only be valid if a correct model has been
found. We emphasize that the saturated model could be extended to inference
for groups with very similar adjustments.

Inference for a group follows the single variable case closely. Firstly,
note that the selection event is completely unchanged by the idea of
testing group significance afterwards as we still apply Lasso for model
selection. Thus, we can still invoke the selection event by conditioning on
$A\mathbf{Y}\leq\mathbf{b}$. Based on \cite{fithian2014optimal}, one can
see that $\big(X_{\tilde{S}}^+\big)_{\tilde{G}}\mathbf{Y}\ \big|\Big(
\big(X_{\tilde{S}\setminus \tilde{G}}\big)^\top
\mathbf{Y},A\mathbf{Y}\leq\mathbf{b}\Big)$ does not depend on
$\boldsymbol{\beta}^{\tilde{S}}_{-\tilde{G}}$ such that there are no more
unknown parameters in our model under the null hypothesis
\eqref{eq:selectednullgroup}. Due to this independence from the nuisance
parameters, we can base the inference on
$\big(X_{\tilde{S}}^+\big)_{\tilde{G}}\mathbf{Y}$ or functions thereof. We
advocate the use of the following test statistic 
\begin{equation*}
\sum_{j\in\tilde{G}}\text{sign}\left(\widehat{\beta}_j\right)\left(X_{\tilde{S}}^+\right)_{j}\mathbf{Y}.
\end{equation*}
In words, it is a directed sum of projections in to all directions
corresponding to the group's variables. Including
$\text{sign}\big(\widehat{\beta}_j\big)$ in our test statistic is valid,
since we additionally condition on having observed the parameters' signs
for the sampling procedure. This additional conditioning is not mandatory
for valid inference but simplifies the computation (cf.\ Section
\ref{datacarving}). The success of the sum can be intuitively justified as
potentially no variable has a significant effect by itself, but the group
as a whole could have. 

There are two main reasons to perform such a group test instead of aggregating p-values of single variables in the group. First, since we are interested in the null hypothesis for the group, it seems more appropriate to conduct a test that treats all the variables within in the group in the same way instead of applying and aggregating multiple tests each of which focussing on a different variable. Second, as the calculation of any null distribution requires to sample a MCMC chain, fewer chains have to be created when looking at a group simultaneously. Though, this comes at the price of a higher dimensionality to sample from compared to treating covariates individually.

We need to sample from the (approximate) null distribution to perform tests. As in the single variable case,
the carving procedure leads to a Gaussian distribution subject to linear equality and
inequality constraints, which can be sampled from as presented in Appendix
\ref{app:mcmc} with few adjustments. 

In contrast to testing of single variables, the group problem remains
multidimensional in the saturated view (for $\big\vert
\tilde{G}\big\vert>1$) as one conditions on all but the group's variables. To
sample from this saturated model, some more changes would be needed,
especially the conditioning in \ref{app:eqcon} has to be adjusted, while
\ref{app:lintrafo} has to be omitted.

In Section \ref{prop:group}, we establish the validity of our group test on a single split. This validity is enough to enable multicarving using standard aggregation techniques \eqref{eq:fixquant} or \eqref{eq:optquant}. When testing for several groups, the fourth step of the multicarving procedure given in Section \ref{mdc} must be adapted to a suitable multiplicity correction factor. The factor which enlarges the p-values naturally depends on the construction of the different groups. Some possible choices for disjoint groups are $p / {\big\vert G \big\vert}$ and ${\big\vert \tilde{S}^{\left[b\right]} \big\vert}/ {{\big\vert \tilde{S}^{\left[b\right]}\cap G \big\vert}}$, where the latter can be different for every split. For a more elaborate description of this procedure as well as an extension to hierarchical testing, see \cite{manpb16} and
\cite{renaux2018hierarchical}.

\subsection{Extension to logistic regression} \label{GLM}
Not all data can be described and approximated well by the linear model given in \eqref{eq:linmod}. 
We extend the inference method to be applicable to generalized linear models 
and focus on logistic regression only in the following. Many of the ideas could be carry over to different generalized linear models too, after applying the right transformations.

In logistic regression, we have a binary response vector $\mathbf{Y}\in \left\{0,1\right\}^n$ and some matrix of predictor variables $X\in \mathbb{R}^{n\times p}$. For every entry $Y_i$ of $\mathbf{Y}$, the probability of being $1$ is modelled as 
\begin{equation}\label{eq:logmod}
\PR{Y_i=1\big|X_i}\ =\ \pi\left(X_i\right)\ =\ \pi_i\ =\ \dfrac{\text{exp}\left(X_i\boldsymbol{\beta}\right)}{1+\text{exp}\left(X_i\boldsymbol{\beta}\right)}
\end{equation}
for some unknown parameter vector $\boldsymbol{\beta}\in \mathbb{R}^p$, the target of our inference. 
We denote the $i$-th row of $X$ by $X_i$.

In a classical low-dimensional setting with $p<n$, this would be fitted using the MLE or equivalently by minimizing the negative of the log-likelihood for an observation $\mathbf{y}$. The log-likelihood $l\left(\boldsymbol{\beta}\right)$ is defined as 
\begin{align*}
l\left(\boldsymbol{\beta}\right)& =\sum_{i=1}^n{\text{log}\left(\PR{Y_i=y_i\big|X_i}\right)}=\sum_{i=1}^n{y_i \text{log}\left(\pi_i\right) + \left(1-y_i\right)\text{log}\left(1-\pi_i\right)}\\
& =\sum_{i=1}^n{y_i X_i\boldsymbol{\beta}-\text{log}\left(1+\text{exp}\left(X_i\boldsymbol{\beta}\right)\right)}.
\end{align*}
The negative of the above formula can be minimized, for example, by using a Newton algorithm, which leads to solving an iteratively reweighted least squares (IRLS) problem as derived in \cite{hastie2009elements}. Starting with some initial estimate $\widehat{\boldsymbol{\beta}}^0$, one iterates 
\begin{equation*}
\widehat{\boldsymbol{\beta}}^{t+1} =\ \left(X^{\top}{W}X\right)^{-1}X^{\top}{W}\mathbf{y}_{adj}\ =\ \underset{\boldsymbol{\beta}}{\text{arg min}}\dfrac{1}{2}\left(\mathbf{y}_{adj}-X\boldsymbol{\beta}\right)^{\top}{W}\left(\mathbf{y}_{adj}-X\boldsymbol{\beta}\right),
\end{equation*}
where we define
\begin{equation*}
{W}=\begin{pmatrix}\widehat{\pi}_1^t\left(1-\widehat{\pi}_1^t\right) && 0 && \cdots && 0\\
0&&\widehat{\pi}_2^t\left(1-\widehat{\pi}_2^t\right)&&\ddots&&0\\
\vdots&&\ddots&&\ddots&&0\\
0&&\cdots&&0&&\widehat{\pi}_n^t\left(1-\widehat{\pi}_n^t\right)
\end{pmatrix}, \quad \mathbf{y}_{adj}=X\widehat{\boldsymbol{\beta}}^t+{W}^{-1}\left(\mathbf{y}-\widehat{\boldsymbol{\pi}}^t\right).
\end{equation*}
Thus, in every step a weighted least-squares problem with weight matrix ${W}$, which iteratively changes, is solved. This explains the name of the procedure.

By further defining 
\begin{equation*}
\mathbf{y}_w=\sqrt{{W}}\mathbf{y}_{adj}, \quad X_w=\sqrt{{W}}X,
\end{equation*}
this can be reformulated as a usual least-squares problem \citep{dezeure2015high}
\begin{equation*}
\widehat{\boldsymbol{\beta}}^{t+1}=\underset{\boldsymbol{\beta}}{\text{arg min}}\dfrac{1}{2}\left(\mathbf{y}_{w}-X_w\boldsymbol{\beta}\right)^{\top}\left(\mathbf{y}_{w}-X_w\boldsymbol{\beta}\right).
\end{equation*}
In the low-dimensional case, \cite{dezeure2015high} suggest to perform the inference as if the final iterate follows $\mathbf{Y}_w \sim \mathcal{N}\left(X_w\boldsymbol{\beta},I\right)$. This approach is asymptotically valid because if this was the case, one would have 
\begin{equation*}
\widehat{\boldsymbol{\beta}}=\left(X_w^{\top}X_w\right)^{-1}X_w^{\top}\mathbf{Y}_w \sim \mathcal{N}\left(\boldsymbol{\beta},\left(X_w^{\top}X_w\right)^{-1}\right),
\end{equation*}
which is the limiting distribution of the MLE. This can be seen by noting that the covariance matrix is the plug-in estimate of the inverse Fisher information.

As for the linear model \eqref{eq:linmod}, the MLE cannot be uniquely found for $p>n$ since $X^{\top}{W}X$ is not invertible anymore. Therefore, one also depends on some sort of shrinkage. One can use the Lasso, i.e.,\ an $\ell_1$-penalty, in the same fashion as for the linear model and solve the following minimization 
\begin{equation*}
\widehat{\boldsymbol{\beta}}=\underset{\boldsymbol{\beta}}{\text{arg min}} \ -l\left(\boldsymbol{\beta}\right)+\lambda\left\Vert\boldsymbol{\beta}\right\Vert_1 .
\end{equation*}
This minimizer can be found similarly as in the non-penalized case by adding the penalty term in every update \citep{friedman2010regularization}
\begin{align*}
\widehat{\boldsymbol{\beta}}^{t+1}& =\underset{\boldsymbol{\beta}}{\text{arg min}}\dfrac{1}{2}\left(\mathbf{y}_{adj}-X\boldsymbol{\beta}\right)^{\top}{W}\left(\mathbf{y}_{adj}-X\boldsymbol{\beta}\right)+\lambda\left\Vert\boldsymbol{\beta}\right\Vert_1\\
& =\underset{\boldsymbol{\beta}}{\text{arg min}}\dfrac{1}{2}\left(\mathbf{y}_{w}-X_w\boldsymbol{\beta}\right)^{\top}\left(\mathbf{y}_{w}-X_w\boldsymbol{\beta}\right)+\lambda\left\Vert\boldsymbol{\beta}\right\Vert_1 .
\end{align*}
Thus, the final Lasso estimate will (approximately) fulfil
\begin{equation*}
\widehat{\boldsymbol{\beta}}=\underset{\boldsymbol{\beta}}{\text{arg min}}\dfrac{1}{2}\left(\mathbf{y}_{w}-X_w\boldsymbol{\beta}\right)^{\top}\left(\mathbf{y}_{w}-X_w\boldsymbol{\beta}\right)+\lambda\left\Vert\boldsymbol{\beta}\right\Vert_1,
\end{equation*}
where $X_w$ and $\mathbf{y}_w$ are functions of the estimate $\widehat{\boldsymbol{\beta}}$ itself. As this is exactly a Lasso fit as in \eqref{eq:lasso}, the estimate $\widehat{\boldsymbol{\beta}}$ will also fulfil the KKT criteria defined by $X_w$ and $\mathbf{y}_w$. Therefore, we can formulate the constraint $A\mathbf{Y}_w\leq\mathbf{b}$, which the observed adjusted response is required to fulfil.

In the high-dimensional case with Lasso selection, it is an obvious approach to calculate inference statements as if $\mathbf{Y}_w \sim \mathcal{N}\left(X_w\boldsymbol{\beta},I\right)\ \big| A\mathbf{Y}_w\leq\mathbf{b}$ inspired by the inference techniques in the low-dimensional setting. Or in other words, proceed as in the usual Gaussian case using our new transformed data $X_w$ and $\mathbf{Y}_w$. This can be done likewise for either pure post-selection inference or data carving.

\cite{taylor2018post} provide an argument for the first case. Their main assumption is $\sqrt{n}$-consistency of the Lasso estimator $\widehat{\boldsymbol{\beta}}$. This condition is discussed, for example, in \cite{buhlmann2011statistics}. Under this assumption, the ``one-step estimator'' $\bar{\boldsymbol{\beta}} \equiv X_{w,\tilde{S}}^+\mathbf{Y}_w$ would have the same limiting Gaussian distribution as the usual low-dimensional MLE if no selection was applied. After some technicalities, which we do not want to recite here, they are able to derive the corresponding constrained limiting distribution from this non-selective CLT.

Importantly, this theory was derived for the fixed-$p$ case. Especially, $\sqrt{n}$-consistency of the Lasso estimator typically only holds for fixed $p$. An argument for the high-dimensional case $p \gg n \rightarrow \infty$, if any exists, is yet to be found. Recent developments by \cite{sur2019modern} and \cite{zhao2020asymptotic} regarding the limiting distribution of the MLE suggest that one has to additionally assume at least $s = { \scriptstyle \mathcal{O}}\left(n\right)$ in order to derive such an argument.

We empirically test the adaption of pure post-selection inference for logistic regression to data carving in our simulations without giving a full theoretical argument. Presumably, such an argument, at least for the fixed-$p$ case, could follow using similar concepts as in \cite{taylor2018post}.

Other types of generalized linear models are often fitted in the same fashion using (penalized) IRLS. Whenever this is the case, one can apply our carving method to the transformed data, i.e.,\ $X_w$ and $\mathbf{y}_w$, which behave asymptotically Gaussian.
\paragraph{Multicarving and aggregation.} As in Section \ref{mdc}, we apply this method of calculating p-values to various splits and aggregate as described in Section \ref{ms}. Those aggregation techniques are proven to be unbiased given screening. Obviously, assuming that aggregation is performed over p-values that are all valid themselves given screening.

Here, the p-values are only asymptotically valid even under
screening. Asymptotic validity of the aggregation over asymptotically valid
p-values has not yet been theoretically studied in depth. Therefore, we
cannot restate the same theoretical results for logistic regression as were
derived in \cite{meinshausen2009p} for multisplitting and which we adapt in
Section \ref{prop:mdc} for multicarving in a linear model. Nevertheless, applying multicarving to logistic regression does not
result in any problem with type-I error control in
our simulations so that we can advocate its
use.

\section{Theoretical properties}
We elaborate here the theoretical properties of multicarving and the
extension to group testing for (multi)carving in the selected view,
requiring the screening assumption in \eqref{ass:tild1}. 
Without the screening assumption, (multi)carving is still valid controlling the type I error in great generality when taking the saturated view. Then, at the price to be often overly conservative, confidence intervals with guaranteed coverage should be preferred over tests, see also Sections \ref{datacarving} and \ref{CI}. Throughout this section, we assume that the data follow the linear model
 \eqref{eq:linmod} with Gaussian errors.

\subsection{Multicarving for the linear model} \label{prop:mdc}
Validity of our multicarve method follows naturally from validity of single-carving and multisplitting. Assuming screening in split $b$ and known variance, we know from the theory of data carving that $p_j^{\left(b\right)}$ as defined in \eqref{eq:carvepv} follows $p_j^{\left(b\right)}\sim \text{Unif}\left[0,1\right]$ for $j\in \tilde{S}$ but $j \notin S$. Basically, this uniformity of the p-value is the only thing needed to construct the proofs of Theorems 3.1 and 3.2 in \cite{meinshausen2009p}. Therefore, we can restate their theoretical result for the aggregation methods. Though, we slightly alter the assumptions on the model selection procedure. We assume
\leqnomode
\begin{flalign}
& \qquad \text{ Asymptotic screening: lim}_{n\rightarrow
 \infty}\PR{\tilde{S}\supseteq S}=1\ \text{(as in Section \ref{ms})}. && \tag{A1} \label{ass:tild1}\\ 
& \qquad \text{ Sparsity: }\tilde{s}< n_1. && \tag{\~A2}\label{ass:tild2}
\end{flalign}
\reqnomode
The difference in the second condition yields from the fact that one has to invert $X_2^\top X_2$ to perform inference using splitting, while $X_1^ \top X_1$ has to be inverted for data carving. Actually, the condition is $\text{rank}\big(X_{1,\tilde{S}}\big)=\tilde{s}$ and we implicitly assume this to follow from the sparsity condition. Our simulations suggest to use $n_1>n_2$, thus this altered sparsity assumption is less restrictive. Using those two conditions, we establish FWER control for our multicarve procedure.
\begin{theo}\label{theo:fixquant}
Let $\mathbf{Y} $ be generated by the linear model \eqref{eq:linmod} with Gaussian errors. Assume that \eqref{ass:tild1} and \eqref{ass:tild2} apply. Let $\alpha,\gamma\in \left(0,1\right]$. Let $P_j^{\left(b\right)}$ be calculated as in Section \ref{mdc} with known $\sigma$ and let $Q_j\left(\gamma\right)$ be the aggregated value according to \eqref{eq:fixquant} with finite $B$. Then, it holds 
\begin{equation*}
\underset{n \rightarrow \infty}{\text{lim sup}}\ \PR{\underset{j \notin S}{\text{min }}Q_j\left(\gamma\right)\leq \alpha}\leq \alpha,
\end{equation*}
where the probability is with respect to the data sample. The statement holds regardless of the $B$ random sample splits.
\end{theo}
The analogue result holds when aggregation is not done with a fixed quantile $\gamma$ but with the optimized quantile and the adequate correction term.
\begin{theo}
Let $\mathbf{Y} $ be generated by the linear model \eqref{eq:linmod} with Gaussian errors. Assume that \eqref{ass:tild1} and \eqref{ass:tild2} apply. Let $\alpha,\gamma_{min}\in \left(0,1\right]$. Let $P_j^{\left(b\right)}$ be calculated as in Section \ref{mdc} with known $\sigma$ and let $P_j$ be the aggregated value according to \eqref{eq:optquant} with finite $B$. Then, it holds 
\begin{equation*}
\underset{n \rightarrow \infty}{\text{lim sup}}\ \PR{\underset{j \notin S}{\text{min }}P_j\leq \alpha}\leq \alpha,
\end{equation*}
where the probability is as in Theorem \ref{theo:fixquant}.
\end{theo}
For proofs, we refer to the appendix of \cite{meinshausen2009p} invoking the fact that $p_j^{\left(b\right)}$ is stochastically larger than $\text{Unif}\left[0,1\right]$ under our assumptions.

Some more technicalities have to be considered for error control in a practical set-up. First, in
order for the uniformity assumption to hold, we depend on a good
convergence of the MCMC approximation. Second, since we refrain from
conditioning on $\left\Vert\mathbf{Y}\right\Vert^2$, we need to know the
variance, which is often rather unrealistic. Though, we emphasize that the
same theoretical result would hold in the unknown variance case when
actually using the conditioning trick. Further, when using an overestimate
of $\sigma$, tests become likely more conservative such that type-I error control
is given at least as good as with the true variance parameter. However,
this cannot be guaranteed in all cases. A discussion on this issue can for
example be found in the supplemental materials of
\cite{tibshirani2018uniform}. Third, since we work with finite data, there is no way to guarantee the screening assumption in general. For analogous reasons as argued in Section \ref{ms}, chances are that multicarving corrects the type-I error better than single-carving in such set-ups. However, if screening becomes too unlikely, breaches in the error rate are likely to happen for multicarving as well. This is especially an issue for highly correlated covariates which make the Lasso selection very difficult. We analyze this effect in our simulations in Section \ref{res:mdc} and Appendix \ref{app:mdc}.

\subsection{Data carving for group testing}\label{prop:group}
In this section, we focus on the theoretical properties of our group test applied to a single group using a single split. Using Theorem \ref{theo:groupunif}, results for multicarving then follow from standard arguments.

At the base of our group test is the following lemma, which is proven in Appendix \ref{app:proofs}.
\begin{lemm} \label{lemm:group}
Let $\mathbf{Y} $ be generated by the linear model \eqref{eq:linmod} with Gaussian errors. Let $G$ be some group with $\big\vert\tilde{G}\big\vert>0$, where $\tilde{G}=G \cap \tilde{S}$. Assume that the screening property ($\tilde{S}\supseteq S$) and \eqref{ass:tild2} hold, and $\sigma$ is known. Then, the probability law of 
\begin{equation*}
\left(X_{\tilde{S}}^+\right)_{\tilde{G}}\mathbf{Y}\ \Big|\left(X_{\tilde{S}\setminus \tilde{G}}\right)^\top \mathbf{Y},A\mathbf{Y}\leq\mathbf{b}
\end{equation*}
is completely defined by our parameter of interest $\boldsymbol{\beta}^{\tilde{S}}_{\tilde{G}}$.
\end{lemm}
Using this lemma, we can base our inference statement on the conditional distribution of $\left(X_{\tilde{S}}^+\right)_{\tilde{G}}\mathbf{Y}$. Let $\mathbf{y}$ be some observation, then we define our selected group p-value as
\begin{align} \label{eq:grouppv}
p_{\tilde{G}}\left(\mathbf{y}\right)=\mathrm{P}\bigg[ \ & \sum_{j\in\tilde{G}}\text{sign}\left(\widehat{\beta}_j\right)\left(X_{\tilde{S}}^+\right)_{j}\mathbf{Y}\geq\sum_{j\in\tilde{G}}\text{sign}\left(\widehat{\beta}_j\right)\left(X_{\tilde{S}}^+\right)_{j}\mathbf{y}\ \bigg| \nonumber\\
& \boldsymbol{\beta}^{\tilde{S}}_{\tilde{G}} =0,\left(X_{\tilde{S}\setminus \tilde{G}}\right)^{\top}\mathbf{Y}=\left(X_{\tilde{S}\setminus \tilde{G}}\right)^{\top}\mathbf{y},A\mathbf{Y}\leq\mathbf{b}\bigg].
\end{align}
This probability can be calculated since we additionally condition on the only remaining unknowns in the model. Notably, this exactly defines the ``probability of observing a value at least as extreme as the observed statistic'' under null hypothesis \eqref{eq:selectednullgroup}. Thus, it fulfils the desired property of a p-value, which leads to the following theorem.
\begin{theo} \label{theo:groupunif}
Let $\mathbf{Y} $ be generated by the linear model \eqref{eq:linmod} with Gaussian errors. Assume that the screening property ($\tilde{S}\supseteq S$) and \eqref{ass:tild2} hold, and $\sigma$ is known. Let $\mathbf{y}$ be a realization of $\mathbf{Y}$ and $p_{\tilde{G}}\left(\mathbf{y}\right)$ for some group $G$ with $\big\vert \tilde{G}\big\vert>0$ be calculated as in \eqref{eq:grouppv}. Then, under null hypothesis \eqref{eq:selectednullgroup}, it holds
\begin{equation*}
p_{\tilde{G}}\left(\mathbf{Y}\right) \sim \text{Unif}\left[0,1\right].
\end{equation*}
\end{theo}
Now further define a general group p-value for group $G$ as
\begin{equation}\label{eq:grouppv2}
p_{G}\left(\mathbf{y}\right)=
\begin{cases}
p_{\tilde{G}}\left(\mathbf{y}\right) & \text{if } \big\vert\tilde{G}\big\vert>0\\
1 & \text{else}.
\end{cases}
\end{equation}
Then, we can establish error control of our procedure.
\begin{theo} \label{theo:grouperror}
Let $\mathbf{Y} $ be generated by the linear model \eqref{eq:linmod} with Gaussian errors. Assume that the screening property $\big(\tilde{S}\supseteq S\big)$ and \eqref{ass:tild2} hold, and $\sigma$ is known. Let $\mathbf{y}$ be a realization of $\mathbf{Y}$ and $p_{G}\left(\mathbf{y}\right)$ for some group $G$ be calculated as in \eqref{eq:grouppv2}. Then, under null hypothesis \eqref{eq:fullnullgroup} and for any $\alpha\in \left(0,1\right]$, it holds
\begin{equation*}
\PR{p_{G}\left(\mathbf{Y}\right)\leq \alpha}\leq \alpha.
\end{equation*}
\end{theo}
The proof is available in the Appendix \ref{app:proofs}. The technicalities mentioned at the end of Section \ref{prop:mdc} apply in the same fashion for our group test.

\section{Numerical results}\label{res}
In this section, we provide detailed results of the performance of our proposed
methods in simulation studies. All results were obtained using the
programming language \textsf{R} \citep{R2020R}.
As an overall summary, we find that multicarving exhibits
often an advantage, sometimes being substantial, over multisplitting or
single-carving methods.

\subsection{Multicarving for the linear model}\label{res:mdc}
We tested our multicarve method testing for single variables in the linear
model on several simulation set-ups and we present here the results for
two of them. In the Appendix \ref{app:mdc}, we add further results for variations of these set-ups where we also show the limitations of (multi)carving.

We do not restrict ourselves to successful screening, we assume the variance to
be unknown and estimate it, and lastly, we select our model through
cross-validated Lasso with regularization parameter $\lambda_{1se}$. All these choices are (in part only slightly) deviating from our theoretical
assumptions. In particular, by choosing $\lambda$ through
cross-validation, more information of $\mathbf{Y}$ is used than invoked in
the selection event, making the inference biased. There are first
approaches to correct for this additional bias, for example, in
\cite{tian2018selective}. However, we refrain from applying any of these,
since they will get computationally more involved and because our empirical
results do not show any significant violation of the selective type-I error 
rate \eqref{eq:selerror} using
cross-validation. Perhaps though, this should be done with a certain
precaution as, e.g.,\ \cite{taylor2018post} report bad error control using a
cross-validated $\lambda$ for post-selection inference in a Cox model. 

We vary the number of splits $B$ in $\{1,10,20,30,40,50\}$ and the fraction $f$
of data used for selection in $\{0.5,0.75,0.9,0.95,0.99,1\}$. In order to keep this section
well-arranged, we restrict ourselves to reporting results for $B=1$ and
$B=50$. Generally, results for different values of $B>1$ are qualitatively
similar with a tendency to get slightly better with increasing
$B$. Naturally, $f=1$ does only make sense for a single split. 

For aggregation over the different splits, we optimize over quantiles as in
\eqref{eq:optquant}. Starting with the default value in the multisplitting
literature, $\gamma_{min}=0.05$, we noticed that this makes the procedure
sometimes overly optimistic leading to poor error control. For some intuition of this effect, assume that there is a true active predictor $\mathbf{X}_j$ and a decently correlated predictor $\mathbf{X}_k$ for which the null hypothesis holds true. In order to falsely reject this null hypothesis, $\mathbf{X}_k$ must be selected as a proxy for $\mathbf{X}_j$ in at least $\gamma_{min} B$ of the random sample splits. Of course, this is more likely the lower we set $\gamma_{min}$. Therefore, we additionally consider 
$\gamma_{min}=0.3$ to have a comparison.
Using a larger $\gamma_{min}$ is
also favorable for computational reasons since less MCMC samples are required 
to be able to find a significant aggregated p-value for the smallest possible 
quantile, namely the $\gamma_{min}$-quantile; see Appendix \ref{app:quant} for more details.

\subsubsection{Toeplitz design}\label{res:toep}
In a first scenario, we sample $X$ once from a multivariate Gaussian
distribution with mean zero and a Toeplitz covariance matrix $\Sigma$ with
$\Sigma_{ij} = \rho^{|i-j|}$ with $\rho=0.6$, and we then treat it as fixed
design. The coefficient vector $\boldsymbol{\beta}$ is
5-sparse, and the active predictors are $\left\{1,5,10,15,20\right\}$, each
of which having a coefficient equal to $1$. The standard deviation is set
to $\sigma=2$, leading to a signal-to-noise ratio (SNR) of approximately
$1.71$. For each simulation run, the variance estimate $\widehat{\sigma}^2$ is calculated through
cross-validated Lasso on the entire data set and is used globally for all
splits and inference methods. 

\begin{figure}[htb!]
  \centering
  \includegraphics[width=1\textwidth]{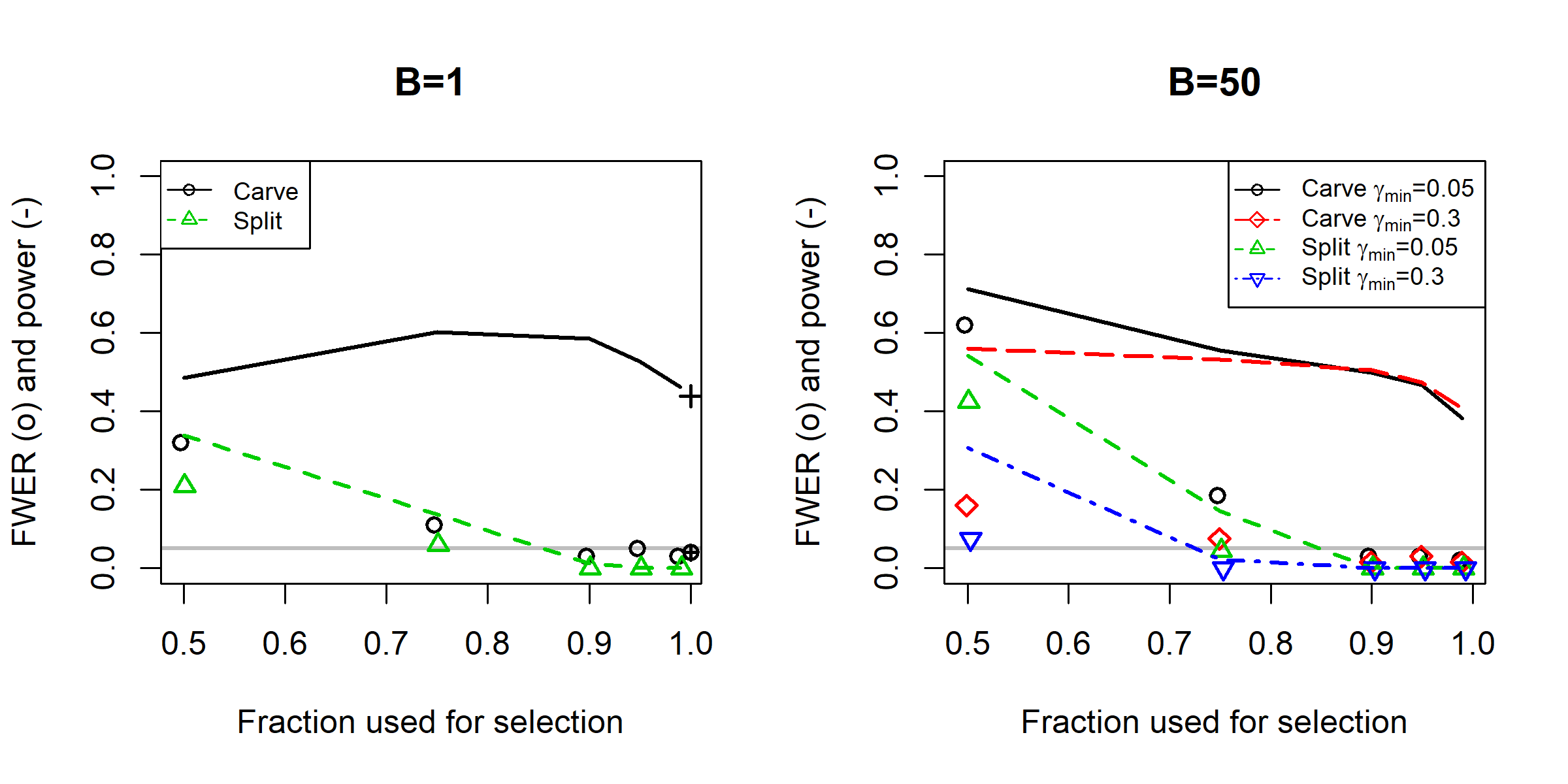}
  \caption[Toeplitz design: unadjusted results]
  {Results for the Toeplitz design. Results using a single split on the left, results using multiple splits on the right. On the x-axis: fraction of data $f$ used for the selection. On the y-axis: FWER depicted by symbols and power depicted by lines. For $f=1$, the power is represented by a cross and the FWER is represented by a circle including a cross. Symbols for the FWER are slightly horizontally offset for better visibility. The horizontal line indicates the target level of the FWER at $\alpha=0.05$. The parameter $\gamma_{min}$ for aggregation is defined in \eqref{eq:optquant}.}
  \label{fig:toepunadj}
\end{figure}

In Figure \ref{fig:toepunadj}, we present the outcome of the simulations for the Toeplitz design. Each performance measure represents $200$ simulation runs. Although screening cannot always be guaranteed, FWER and power are calculated with respect to \eqref{eq:fullnull} with rejection level $\alpha=0.05$. Carving using the entire data for selection, i.e.,\ $f=1$, is performed using 
a different algorithm, namely, the exact calculation from \cite{lee2016exact}. 
We emphasis this using a cross in the figures.
\begin{figure}[!htb]
  \centering
  \includegraphics[width=1\textwidth]{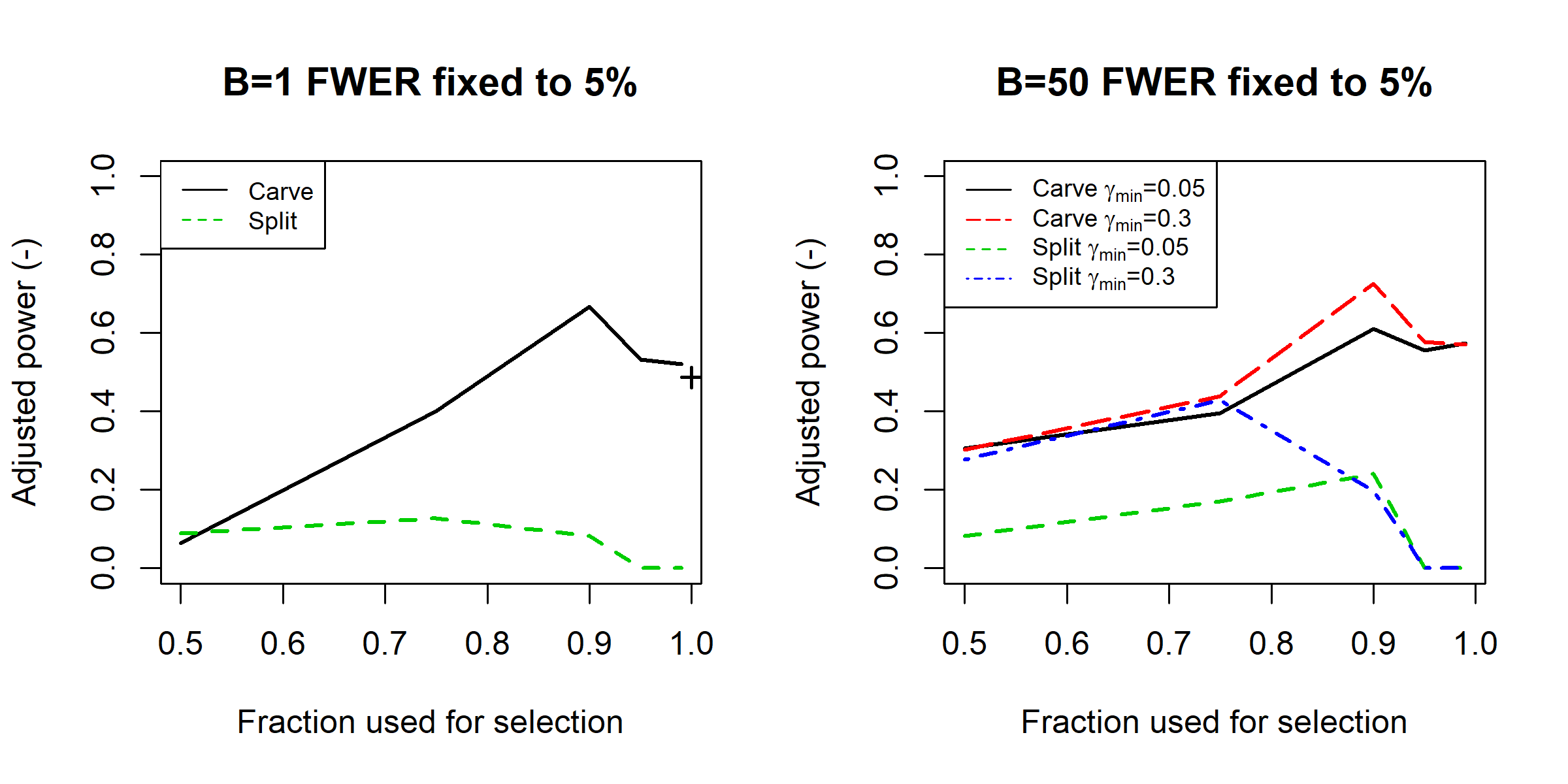}
  \caption[Toeplitz design: adjusted results]
  {Results for the Toeplitz design for the adjusted power. Results using a single split on the left, results using multiple splits on the right. On the x-axis: fraction of data $f$ used for the selection. On the y-axis: adjusted power such that all methods have FWER of exactly $5\%$. For $f=1$, the power is represented by a cross. The parameter $\gamma_{min}$ for aggregation is defined in \eqref{eq:optquant}.}
  \label{fig:toepadj}
\end{figure}

The left-hand side of Figure \ref{fig:toepunadj} illustrates that neither single-splitting nor single-carving controls the error at $5\%$ for $f=0.5$ and $f=0.75$. Though, this is not a violation of our theoretical result, error control would hold when only looking at successful screening. For carving, the power initially increases in $f$ and decreases in the larger values of $f$. This can be explained by the trade-off between more successful screening of the true active set and losing power for the inference stage as more constraints are imposed. The same holds for splitting and multicarving when looking at lower values of $f$ as eventually too few active variables are selected in the first stage such that no decent power remains. As indicated by the inadmissibility statement in \cite{fithian2014optimal}, carving outperforms splitting with respect to power. The important question is now whether multicarving introduces some improvement over single-carving. The single-carve method has the highest power starting from $f=0.75$, where $f=0.5$ can be basically ignored as error control is not given at all. The multicarve method with $\gamma_{min}=0.3$ performs best among all carving methods regarding FWER for all values of $f$. Multicarving with $\gamma_{min}=0.05$ seems to be inferior in this scenario. Thus, there is a trade-off between higher power and better error control. The highest power with $\text{FWER}\leq 5\%$ is obtained at $f=0.9$ for all carving methods with a value of $0.59$ (single-carving), $0.51$ ($\gamma_{min}=0.3$) and $0.50$ ($\gamma_{min}=0.05$). So, the single-carve method is favorable in this situation.

However, this comparison is not quite fair since the methods have different FWER. Therefore, we additionally look at an adjusted power, i.e.,\ the rejection level of the underlying hypothesis tests is adjusted such that each method has an FWER of exactly $5\%$ for each value of $f$; see Figure \ref{fig:toepadj}.
Carving is still superior to splitting although the multisplit method with $\gamma_{min}=0.3$ is now competitive for lower values of $f$. All three carving methods reach their optimum at $f=0.9$, with an adjusted
power of $0.67$ (single-carving), $0.73$ ($\gamma_{min}=0.3$) and $0.61$
($\gamma_{min}=0.05$).

In Appendix \ref{app:toep}, we present further results for Toeplitz designs where $\rho$ is changed to $0.3$ and $0.9$ respectively. Our assumption that the correlation level highly affects the performance is confirmed. Especially, none of the methods in scope performs particularly well for the scenario with $\rho=0.9$ since the initial selection using the Lasso is very unlikely to screen successfully.
\paragraph*{Saturated viewpoint.} 
As discussed in Section \ref{CI}, testing for null hypotheses \eqref{eq:selectednull} (single-carving) and \eqref{eq:multiplenull} (multicarving) while omitting the screening assumption is not particularly meaningful as $\boldsymbol{\beta}^{\tilde{S}^{\left(b\right)}}$ is fully dense. Therefore, the saturated viewpoint without the screening assumption has no advantage for testing null hypotheses. However, in order to assess the power drop mentioned in Section \ref{datacarving}, we test for null hypothesis \eqref{eq:fullnull} using inference in the saturated model. For the set-up discussed above, this leads to the following performance measures. The highest power with $\text{FWER}\leq 5\%$ is $0.44$ for single-carving ($f=1$), $0.44$ for multicarving with $\gamma_{min}=0.05$ ($f=0.9$) and $0.41$ for multicarving with $\gamma_{min}=0.3$ ($f=0.95$). The corresponding highest adjusted power is $0.50$ (single-carving), $0.61$ ($\gamma_{min}=0.05$) and $0.71$ ($\gamma_{min}=0.3$), all of which obtained at $f=0.9$. Thus, the saturated approach leads to lower power and adjusted power as anticipated. Though, this drop is less distinct for the adjusted power as the additional conservatism also leads to better type-I error control. Furthermore, we see that for multicarving the differences are less pronounced than for single-carving. For computational reasons, the saturated viewpoint might, therefore, be an interesting alternative for our multicarve procedure.

In Section \ref{CI}, we further introduced the idea of multicarving confidence intervals, where omitting the screening assumption and using the saturated method appears to be more natural. We present a corresponding analysis in Section \ref{res:CI}.

\paragraph*{PoSI.} In Section \ref{intro}, we mentioned the work by \cite{kuchibhotla2020valid} which generally provides stronger guarantees at the price of increased conservatism. In order to assess this conservatism, we executed a small simulation study applying their method to this Toeplitz design. For this, we use the software available in the GitHub repository cited in \cite{kuchibhotla2020valid}.

For the model selection, we use cross-validated Lasso on all data. Thus, the models on which we perform inference are the same as for pure post-selection inference used above. After calculating the $95\%$ confidence regions for $\boldsymbol{\beta}^{\tilde{S}}$ in all ${\tilde{s}}$ dimensions, we reject the null-hypothesis for covariates $j$ for which $0$ is not within the region. With this technique, we did not receive a single rejection over $1000$ simulation runs. Thus, the expectation that the inference method is very conservative is confirmed.

Since their method is not restricted to Lasso selection but allows for any possible method in the selection step, we tried a different approach. Namely, we applied an ``oracle'' selection that always selects the correct submodel, i.e.,\ $\tilde{S}=S$. However, not a single rejection was observed even using this best possible selection. This further confirms the assumption that guaranteeing simultaneous coverage in all submodels is too restrictive for this simulation set-up.

\subsubsection{Semi-synthetic Riboflavin data} \label{res:ribo}
Since simulated data sometimes behaves somewhat more nicely than real data, we also test the methods on ``semi-synthetic'' set-ups, meaning that the $X$ matrix comes from some real data set. We simulate the response $\mathbf{Y}$ from \eqref{eq:linmod} with known $\boldsymbol{\beta}$.

We use the Riboflavin data set with $n = 71$ and $p = 4088$, which was made publicly available by \cite{buhlmann2014high}. The original response measures the Riboflavin production rate for $71$ samples of strains of Bacillus subtilis and gives the data its name. The $X$ matrix contains the log-expression level of $4088$ genes for each of these strains.

For our simulations, we set $\boldsymbol{\beta}$ to be $2$-sparse and use 
an SNR of $16$. The active variables are chosen at random for every simulation 
run and their respective coefficient is set to 1. 
Since this can result in very different signal strength depending on the 
correlation between the 2 variables, we fix the SNR on a per run basis by always 
adjusting $\sigma$ such that $\tfrac{\widehat{\text{Var}}\left(X\boldsymbol{\beta}\right)}{\sigma^2}=16$. 
Here, $\widehat{\text{Var}}\left(X\boldsymbol{\beta}\right)$ denotes the 
empirical variance of the true underlying signal. 
We choose this rather sparse set-up with high SNR since otherwise Lasso 
selection works very poorly in this high-dimensional set-up and none of 
the inference methods has good performance. To illustrate this, we repeat 
the same simulation with $4$ active predictors; compare with Appendix 
\ref{app:ribo4}. 

For the selection, we again perform cross-validation on the given split. To be more realistic, we stick to the unknown $\sigma$ assumption. With the estimation technique described before, we realized that $\mathbb{P}\left[\widehat{\sigma}\geq\sigma\right]$ is empirically quite low in this scenario. Therefore, we choose the more conservative approach of calculating a new 
$\widehat{\sigma}$ for every split as
\begin{equation} \label{eq:sigmamodwise}
\widehat{\sigma}^b=\sqrt{\dfrac{\big\Vert\mathbf{y}-X\widehat{\boldsymbol{\beta}}^b\big\Vert^2}{n-\tilde{s}}},
\end{equation}
where $\widehat{\boldsymbol{\beta}}^b$ is calculated on the selection data only but $\mathbf{y}$ and $X$ are the full data.
\begin{figure}[!htb]
  \centering
  \includegraphics[width=1\textwidth]{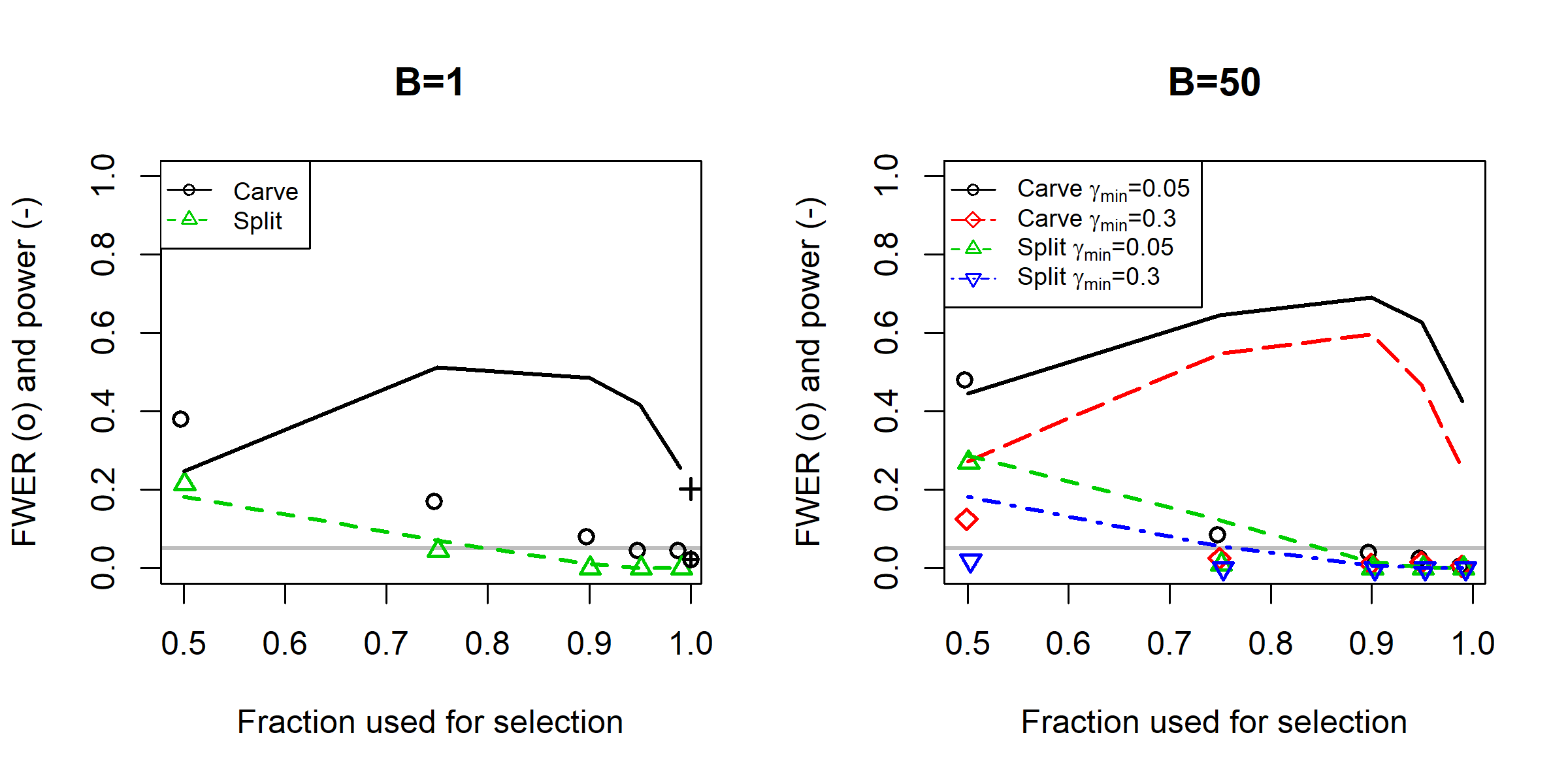}
  \caption[Riboflavin $X$ with sparsity 2: unadjusted results]
  {Results for the Riboflavin $X$ with sparsity 2. See caption of Figure \ref{fig:toepunadj}.}
  \label{fig:ribo2unadj}
    \includegraphics[width=1\textwidth]{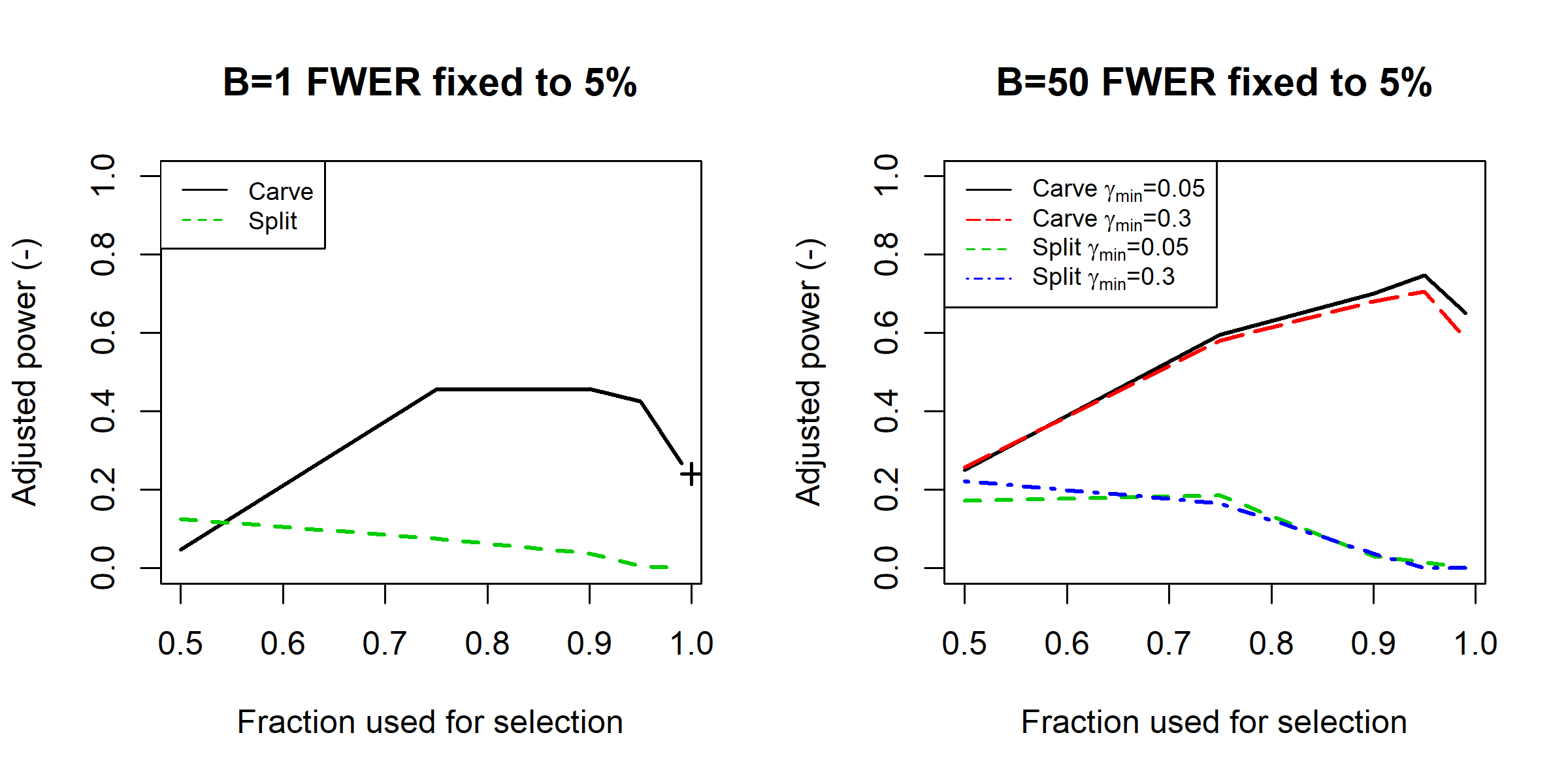}
  \caption[Riboflavin $X$ with sparsity 2: adjusted results]
  {Results for the Riboflavin $X$ with sparsity 2 for the adjusted power. See caption of Figure \ref{fig:toepadj}.}
  \label{fig:ribo2adj}
\end{figure}

The results obtained for the Riboflavin data with a sparsity of $2$ are shown in Figures \ref{fig:ribo2unadj} (FWER and power) and \ref{fig:ribo2adj} (adjusted power). This set-up is now highly in favor of our multicarve method. Especially, the highest power obtained for $\text{FWER}\leq 5\%$ is $0.42$ (single-carving), $0.60$ ($\gamma_{min}=0.3$) and $0.69$ ($\gamma_{min}=0.05$); see Figure \ref{fig:ribo2unadj}. The multicarve methods reach this maximum at $f=0.9$, while single-carving only obtains error control starting from $f=0.95$ and higher. There is a power versus FWER trade-off between the two different values of $\gamma_{min}$. 

The adjusted power is slightly in favor of the lower value $\gamma_{min}=0.05$ as illustrated in Figure \ref{fig:ribo2adj}. More precisely, the highest adjusted power is $0.75$ ($\gamma_{min}=0.05$) and $0.71$ ($\gamma_{min}=0.3$) for the multicarve method. Both these values are obtained for $f=0.95$. Single-carving reaches its maximum of $0.46$ at $f=0.9$. Thus, the adjusted power clearly prefers multicarving as well.

We note that although we increase both SNR and sparsity, the adjusted power is not (much) better than in the previous set-up. This can be intuitively explained by the following two reasons: First, $\tfrac{p}{n} \approx 58$ in the Riboflavin design is much larger than $\tfrac{p}{n} = 2$ in our Toeplitz design. Second, there are variables with a very high empirical correlation of up to around $99\%$, making them hardly distinguishable in the selection stage.

\subsection{Confidence intervals}\label{res:CI}
We apply our method for confidence intervals to the same set-up with $X$ simulated from a multivariate normal distribution with Toeplitz $\rho=0.6$ covariance matrix as in Section \ref{res:toep}. As we explicitly omit the screening assumption, we use a different estimate $\widehat{\sigma}$ for every split as in \eqref{eq:sigmamodwise}. The target parameters $\beta_{j}^{\tilde{S}^{\left(b\right)}}$ in \eqref{eq:cicoverage} are defined including an intercept. Naturally, whenever screening works, this intercept term vanishes.

We use $B=50$ splits and aggregate according to \eqref{eq:optquant} with $\gamma_{min}=0.05$. The obtained intervals are targeted to be $95\%$-confidence intervals ($95\%$-CI).

In Tables \ref{tab:ci} and \ref{tab:cimedquant}, we compare the performance of our carving confidence intervals to the ones obtained using multisplitting implemented in \cite{dezeure2015high}. Those results are based on $200$ simulation runs. 

\begin{table}[b!]
\centering
\begin{tabular}{|lr|ccccc|P{2.5cm}P{2.5cm}|}
\hline
&& \multicolumn{5}{c|}{Median interval length for $95\%$-CI}&\multicolumn{2}{c|}{Average false coverage}\\
&& \multicolumn{5}{c|}{number of active variables}&\multicolumn{2}{c|}{among}\\
 $f$&Method& 1& 5& 10& 15& 20&all $\left[10^{-3}\right]$&tested $\left[10^{-3}\right]$\\
\hline
\multirow{2}{1em}{$0.5$}&Splitting&$1.76$&$1.98$&$1.92$&$1.79$&$1.69$&$2.93$&$6.31$\\
&Carving&$1.88$&$2.41$&$2.14$&$1.95$&$1.85$&$1.6$&$3.47$\\
\hline
\multirow{2}{1em}{$0.75$}&Splitting&$2.42$&$2.77$&$2.7$&$2.44$&$2.3$&$0.23$&$1.07$\\
&Carving&$1.70$&$2.18$&$1.95$&$1.75$&$1.63$&$0.68$&$2.65$\\
\hline
\multirow{2}{1em}{$0.9$}&Splitting&$27.16$&$28.03$&$27.15$&$23.32$&$21.12$&$0$&$0$\\
&Carving&$1.64$&$2.10$&$1.99$&$1.72$&$1.66$&$0.18$&$1.17$\\
\hline
\multirow{2}{1em}{$0.95$}&Splitting&-&-&-&-&-&-&-\\
&Carving&$1.68$&$2.31$&$2.05$&$1.73$&$1.64$&$0.18$&$1.41$\\
\hline
\multirow{2}{1em}{$0.99$}&Splitting&-&-&-&-&-&-&-\\
&Carving&$2.97$&$2.78$&$2.37$&$1.77$&$1.70$&$0.15$&$1.36$\\
\hline
\end{tabular}
\caption[Confidence Intervals]{\label{tab:ci}
Median length for active variables and average false coverage rate of the confidence intervals. The left-hand side displays the median interval length obtained for the true active predictors, i.e.,\ $\left\{1,5,10,15,20\right\}$. The average false coverage rate of the obtained confidence intervals is shown on the right-hand side. This rate is calculated either with respect to all $p=200$ variables or only with respect to variables that are actually tested for, i.e.,\ variables that are at least selected once within the $50$ splits. In this analysis, variables not selected at all are assigned an infinite confidence interval such that no false coverage can occur.
}
\end{table}

The obtained intervals are generally rather conservative as the false coverage rate is always far below the theoretical bound of $5\%$. Notably, for $f=0.5$, the intervals obtained through carving are not actually shorter than those from splitting. The advantage of carving is that the intervals get shorter in a first phase when increasing $f$. By increasing $f$, the selected models become more stable and likewise, $\beta_{j}^{\tilde{S}^{\left(b\right)}}$ differs less over different splits, $b=1, \ldots ,50$. Due to more stable $\beta_{j}^{\tilde{S}^{\left(b\right)}}$, shorter intervals are theoretically possible with higher $f$. Though, multisplitting cannot profit from this as too little information for the inference stage remains after increasing $f$. The same holds for carving when $f$ becomes too large. The best performing method is carving with a selection fraction of $f=0.9$ which outperforms every other configuration with respect to at least three interval lengths. Further, it also performs comparably well with respect to the false coverage rate as every configuration with lower false coverage rate suffers from substantially longer intervals.

In a further analysis, we look at the length of the confidence intervals of all covariates that were selected at least once within the $B=50$ splits. For the other variables, there is no real interpretation of the coverage in Equation \eqref{eq:cicoverage}. Further, not selecting a covariate at all in $50$ splits is a rather strong indication for the variable generally being inactive such that treating it as if it has an infinite confidence interval length does not seem appropriate. However, there are still many variables obtaining an infinite interval length, namely, those that are selected at least once but less than $\gamma_{min}B$ times, i.e.,\ once or twice in this set-up.

In Table \ref{tab:cimedquant}, we report the median over the $200$ simulation runs over several quantiles of the interval lengths among the selected variables. Due to the possibility of infinite interval lengths, we focus on quantiles instead of averages.

\begin{table}[h!]
\centering
\begin{tabular}{|lr|ccccc|}
\hline
&& \multicolumn{5}{c|}{Quantile}\\
 $f$&Method&$10\%$&$20\%$&$30\%$&$40\%$&$50\%$\\
\hline
\multirow{2}{1em}{$0.5$}&Splitting&$2.37$&$3.09$&$4.26$&$\infty$&$\infty$\\
&Carving&$4.49$&$12.46$&$44.59$&$\infty$&$\infty$\\
\hline
\multirow{2}{1em}{$0.75$}&Splitting&$3.1$&$4.22$&$5.86$&$12.19$&$\infty$\\
&Carving&$2.63$&$5.05$&$10.53$&$26.79$&$\infty$\\
\hline
\multirow{2}{1em}{$0.9$}&Splitting&$31.54$&$905.75$&$\infty$&$\infty$&$\infty$\\
&Carving&$2.08$&$3.25$&$4.99$&$9.47$&$19.33$\\
\hline
\multirow{2}{1em}{$0.95$}&Splitting&-&-&-&-&-\\
&Carving&$1.89$&$2.78$&$3.84$&$5.36$&$8.11$\\
\hline
\multirow{2}{1em}{$0.99$}&Splitting&-&-&-&-&-\\
&Carving&$1.94$&$2.53$&$3.44$&$4.69$&$6.13$\\
\hline
\end{tabular}
\caption[Confidence Intervals]{\label{tab:cimedquant}Results of length of confidence intervals. Median is taken over simulation runs of several quantiles over lengths of $95\%$-CI of variables that were selected at least once in $B=50$ splits.}
\end{table}
Again, we note that for $f=0.5$, the intervals obtained through multicarving are longer than those from multisplitting. However, the power of multicarving comes from the ability to raise the selection fraction without losing all information for the inference stage. 
The $50$ selected models become more stable for larger values of $f$ and fewer 
covariates are selected in total over the $B$ splits. 
The total number of distinct variables selected over all the splits is 96 and 
20 on average for $f = 0.5$ respectively $f = 0.99$. With fewer features under consideration, a higher fraction of those is selected sufficiently often such that powerful inference is possible. Those effects are visible in Table \ref{tab:cimedquant} as the quantiles of the intervals using multicarving mostly become shorter when increasing $f$. For carving, there is also a natural countereffect as information for the inference stage is lost, thus the quantiles of interval lengths are not strictly decreasing.

In summary, our confidence intervals obtain the desired coverage stated in Equation \eqref{eq:cicoverage}. Further, multicarving brings an advantage compared to multisplitting because of the possibility to perform well using a higher selection fraction $f$.

\subsection{Data carving for group testing}\label{res:group}
In order to see how well our group test performs, we compare it with results 
presented in \cite{guo2019group}. The authors consider two scenarios testing 
either a small or large group
based on data simulated using different covariance structures. Testing a large 
group in a dense scenario is described below. Results of group testing for a 
small group in a sparse and high correlation scenario are illustrated in the 
Appendix \ref{app:group}. 

The dense alternative with many small non-zero coefficients is a set-up where testing single variables is difficult. More precisely, $p$ is $500$ and $n$ is varied in $\left\{250,350,500,800\right\}$. The feature matrix $X$ is generated from normally distributed features having a Toeplitz covariance matrix with $\rho=0.6$. The parameter vector is defined as $\beta_j=\delta$ for $25\leq j \leq 50$ and $\beta_j=0$ otherwise. We vary $\delta$ over $\left\{0,0.02,0.04,0.06\right\}$ where $\delta = 0$ corresponds to the global null. The response vector $\mathbf{Y}$ follows our linear model \eqref{eq:linmod} with $\sigma=1$. This leads to SNR in $\left\{0,0.039,0.154,0.347\right\}$. We are interested in testing null hypothesis \eqref{eq:fullnullgroup} for the group $G=\left\{30,31, \ldots ,200\right\}$.

\subsubsection{Single-carving for group testing}
We perform inference using our group test introduced in Section \ref{group}. As in Section \ref{res:mdc}, we vary the fraction of data used for selection $f$ in $\left\{0.5,0.75,0.9,0.95,0.99,1\right\}$. We start with just using a single split, i.e.,\ $B=1$, for inference.
Notably, for the group test, inference using $f=1$ is obtained with MCMC sampling as well. Since we condition on all but the covariates of interest, we generally have more than $1$ degree of freedom such that an easy calculation as in \cite{lee2016exact} is not possible. The only exception to that is if $\big\vert \tilde{G}\big\vert=1$, which is algorithmically equivalent to single variable testing.

For the selection, we perform cross-validation. Based on the assumption that Lasso might eliminate many of the covariates with weak signal, we use $\lambda_{min}$ instead of $\lambda_{1se}$. To assess the variance parameter $\sigma$, we use a global estimate obtained with cross-validation and $\lambda_{min}$ on all data.

In Table \ref{tab:dense}, we show the results for the dense alternative. For each combination of $\delta$, $n$, and $f$, we report the empirical rejection rate (ERR), i.e.,\ the fraction out of 200 simulation runs in which the null hypothesis is rejected at level $\alpha=5\%$. For $\delta=0$, this measures the type-I error, for $\delta>0$, this measures the power.
\begin{table}[t]
\centering
\begin{tabular}{|lr|cccccc|}
\hline
$\delta$&$n$&$f=0.5$&$f=0.75$&$f=0.9$&$f=0.95$&$f=0.99$&$f=1$\\
\hline
\multirow{4}{1em}{$0$} &$250$&$0.04$&$0.075$&$0.035$&$0.045$&$0.05$&$0.075$\\
&$350$&$0.025$&$0.045$&$0.08$&$0.055$&$0.035$&$0.05$\\
&$500$&$0.025$&$0.045$&$0.08$&$0.045$&$0.04$&$0.035$\\
&$800$&$0.025$&$0.03$&$0.045$&$0.05$&$0.03$&$0.03$\\
\hline
\multirow{4}{1em}{$0.02$} &$250$&$0.1$&$0.155$&$0.16$&$0.12$&$0.18$&$0.145$\\
&$350$&$0.105$&$0.185$&$0.175$&$0.205$&$0.18$&$0.19$\\
&$500$&$0.16$&$0.23$&$0.255$&$0.275$&$0.25$&$0.24$\\
&$800$&$0.455$&$0.485$&$0.55$&$0.455$&$0.5$&$0.485$\\
\hline
\multirow{4}{1em}{$0.04$} &$250$&$0.46$&$0.6$&$0.62$&$0.59$&$0.68$&$0.6$\\
&$350$&$0.66$&$0.795$&$0.845$&$0.82$&$0.815$&$0.74$\\
&$500$&$0.88$&$0.945$&$0.96$&$0.965$&$0.97$&$0.93$\\
&$800$&$0.98$&$0.995$&$1$&$1$&$1$&$0.995$\\
\hline
\multirow{4}{1em}{$0.06$} &$250$&$0.88$&$0.935$&$0.97$&$0.955$&$0.955$&$0.925$\\
&$350$&$0.96$&$1$&$0.995$&$1$&$1$&$0.985$\\
&$500$&$0.995$&$1$&$1$&$1$&$1$&$0.995$\\
&$800$&$1$&$1$&$1$&$1$&$1$&$1$\\
\hline
\end{tabular}
\caption[Results for the dense alternative group scenario]{\label{tab:dense}Empirical rejection rate at level $5\%$ for the dense alternative using single-carving.}
\end{table}

For fixed $\delta>0$ and $f$, the power increases in the number of observations $n$, and for fixed $n$ and $f$, it increases in the signal strength $\delta$. This conclusion is to be expected.

The FWER is controlled for all combinations of $f$ and $n$, for most combinations even conservatively. The fraction $f=0.5$ has always the lowest power because selection works not overly well. In many settings, $f=1$ is also suboptimal with respect to power as too little power remains for the inference stage. Fractions $f=0.9$ to $f=0.99$ are all competitive and perform similar. This is in good accordance with our results testing for single variables in Section \ref{res:mdc}.

Table \ref{tab:dense} can be compared to \citet[Table 1]{guo2019group} for $\delta$ in $\left\{0,0.04,0.06\right\}$ and $n$ in $\left\{250, 300, 500\right\}$, where six different methods are evaluated in this scenario. Our method with fractions between $f=0.75$ and $f=0.99$ is amongst the best with respect to power in each set-up. Especially, it has clearly higher power than their method $\phi_{\Sigma}\left(1\right)$ for $\delta=0.04$, whereas the power is similar for $\delta = 0.06$. The power of the method $\phi_{\Sigma}\left(0.5\right)$ is comparable to the power of our group test but their method attains slightly lower values. Though, their method $\phi_\Sigma$ controls the error more conservatively such that a clear statement in favor of either method is not possible.

If we summarize the results from the dense scenario in this section and the results from the sparse scenario in Appendix \ref{app:group}, we can state that our method does not have the best performance in all possible set-ups. Though, it is competitive in all of them, while all competitors have some set-ups where they do not work well at all. Thus, our group test, which results from a very simple adjustment of the data carving idea, offers some valuable results.

\subsubsection{Multicarving for group testing}
In Section \ref{res:mdc}, we see that the multicarve method usually has better error control than single-carving. Based on this observation, it is to be expected that multicarving could further improve on group inference in scenarios where the error is not controlled conservatively (cf.\ Table \ref{tab:dense}). Therefore, we test multicarving for group testing as well. Indeed, with multicarving, no ERR above the target level $5\%$ occurs for $\delta=0$ in either alternative. However, the ERR for $\delta>0$, i.e.,\ the power, is sensitive to the choice of the tuning parameters $f$, $\gamma$ or $\gamma_{min}$, and $B$. Especially, in the two scenarios under consideration, aggregation using a fixed quantile clearly outperforms the use of an optimized quantile according to Equation \eqref{eq:optquant}.

In the following, we present results obtained using $B=20$ splits and a fixed quantile for aggregation of $\gamma=0.05$ in order to show the possibilities of multicarving. We emphasize that these choices work comparably well such that in general, when no such comparison is possible, one could expect slightly lower power using multicarving for group testing. Those results are shown in Table \ref{tab:densemc}.
\begin{table}[!htb]
\centering
\begin{tabular}{|lr|ccccc|}
\hline
$\delta$&$n$&$f=0.5$&$f=0.75$&$f=0.9$&$f=0.95$&$f=0.99$\\
\hline
\multirow{4}{1em}{$0$} &$250$&$0.045$&$0.04$&$0.045$&$0.035$&$0.03$\\
&$350$&$0.04$&$0.05$&$0.04$&$0.03$&$0.03$\\
&$500$&$0.025$&$0.03$&$0.05$&$0.02$&$0.02$\\
&$800$&$0.005$&$0.03$&$0.03$&$0.035$&$0.02$\\
\hline
\multirow{4}{1em}{$0.02$} &$250$&$0.16$&$0.15$&$0.13$&$0.12$&$0.12$\\
&$350$&$0.28$&$0.185$&$0.165$&$0.155$&$0.12$\\
&$500$&$0.31$&$0.24$&$0.225$&$0.215$&$0.195$\\
&$800$&$0.61$&$0.515$&$0.55$&$0.435$&$0.42$\\
\hline
\multirow{4}{1em}{$0.04$} &$250$&$0.75$&$0.765$&$0.65$&$0.62$&$0.615$\\
&$350$&$0.885$&$0.865$&$0.89$&$0.83$&$0.83$\\
&$500$&$1$&$0.97$&$0.98$&$0.96$&$0.945$\\
&$800$&$1$&$1$&$1$&$1$&$1$\\
\hline
\multirow{4}{1em}{$0.06$} &$250$&$0.985$&$0.99$&$0.975$&$0.97$&$0.975$\\
&$350$&$1$&$1$&$1$&$1$&$1$\\
&$500$&$1$&$1$&$1$&$1$&$1$\\
&$800$&$1$&$1$&$1$&$1$&$1$\\
\hline
\end{tabular}
\caption[Results for the dense alternative group scenario]{\label{tab:densemc}Empirical rejection rate at level $5\%$ for the dense alternative using multicarving.}
\end{table}

We consider the dense alternative. For multicarving, the highest ERR for $\delta=0$ is $5\%$, whereas it is $8\%$ for single-carving. Naturally, there is some fluctuation involved in those empirical values. Nevertheless, this difference indicates an improvement of multicarving over single-carving. For most scenarios with $\delta>0$, a selection fraction of $f=0.5$ is favorable. The intuitive explanation is that although ${{\big\vert \tilde{S}^{\left[b\right]}\cap G \big\vert}}$ might on average be smaller than with higher fractions $f$, it is still ``big enough'' in a decent number of splits. In these splits, the lower $f$ allows for a more powerful inference statement making the method more powerful overall after aggregation. Notably, using $B=20$ and $\gamma=0.05$ (fixed quantile for aggregation) is equivalent to a Bonferroni corrected minimum p-value (cf.\ Equation \eqref{eq:fixquant}). Thus, only the most significant split is of importance. We now compare the power in Table \ref{tab:densemc} to that for single-carving in Table \ref{tab:dense}. Using a selection fraction of $0.5$, multicarving outperforms any single-carving configuration in all scenarios unless $\delta=0.02$ and $n=250$. Thus, using multiple splits and aggregating can bring a clear improvement. Though, this is rather sensitive to the choice of the tuning parameters as mentioned above.

In summary, the natural extension of our group test using multiple splits leads to a performance boost. Especially, the error can be controlled on a more conservative level using multiple splits. A drawback of the method is its sensitivity to tuning parameters. If those happen to be chosen poorly, power might be lower compared to single-carving.

\subsection{Multicarving for logistic regression}
\begin{figure}[t!]
  \centering
  \includegraphics[width=1\textwidth]{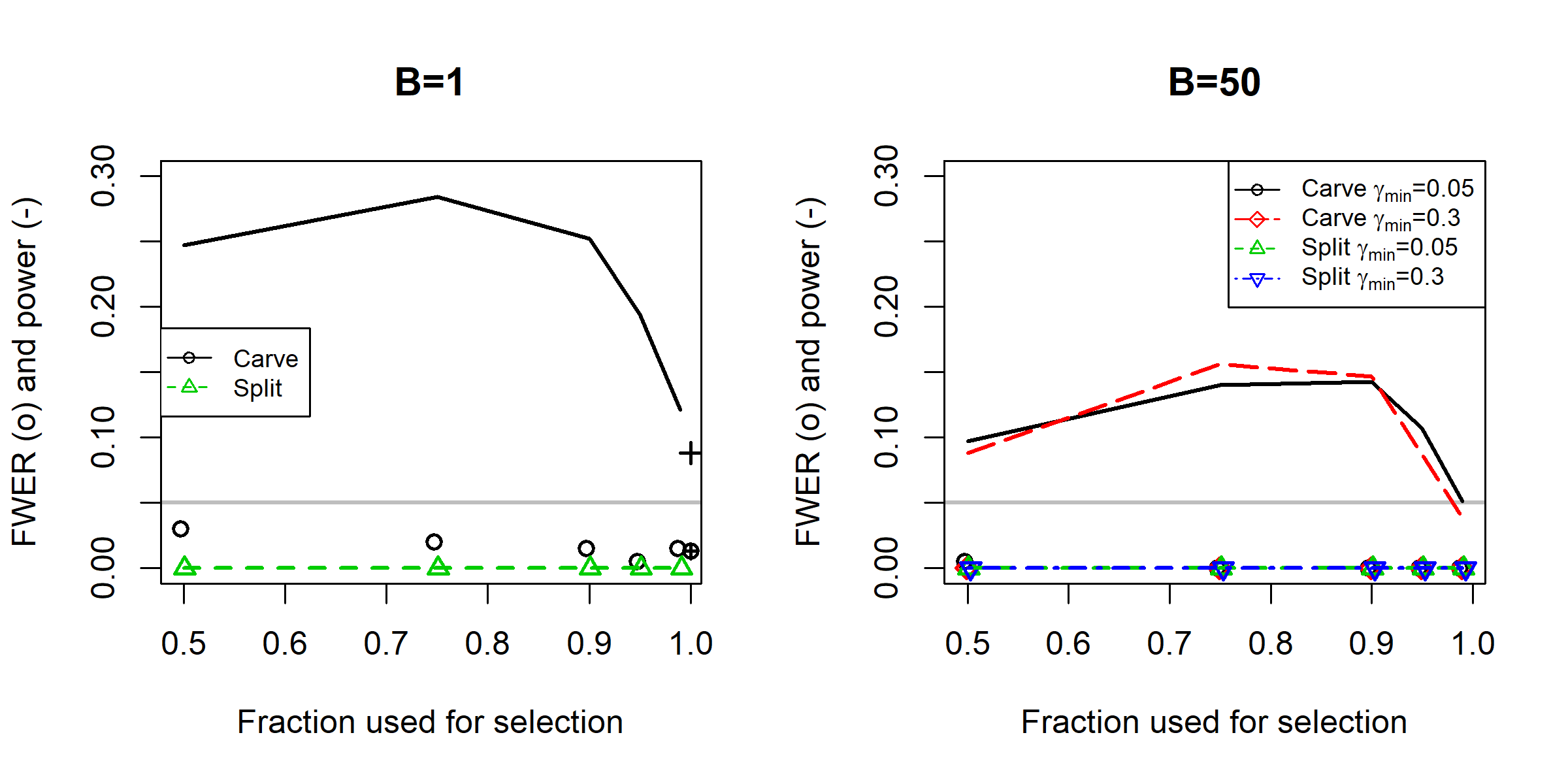}
  \caption[Results for the Toeplitz $X$ in logistic regression.: unadjusted results]
  {Results for the Toeplitz design in logistic regression. See caption of Figure \ref{fig:toepunadj}. Note that the range of values on the y-axis is different compared to all the other figures.}
    \label{fig:binunadj}
    \includegraphics[width=1\textwidth]{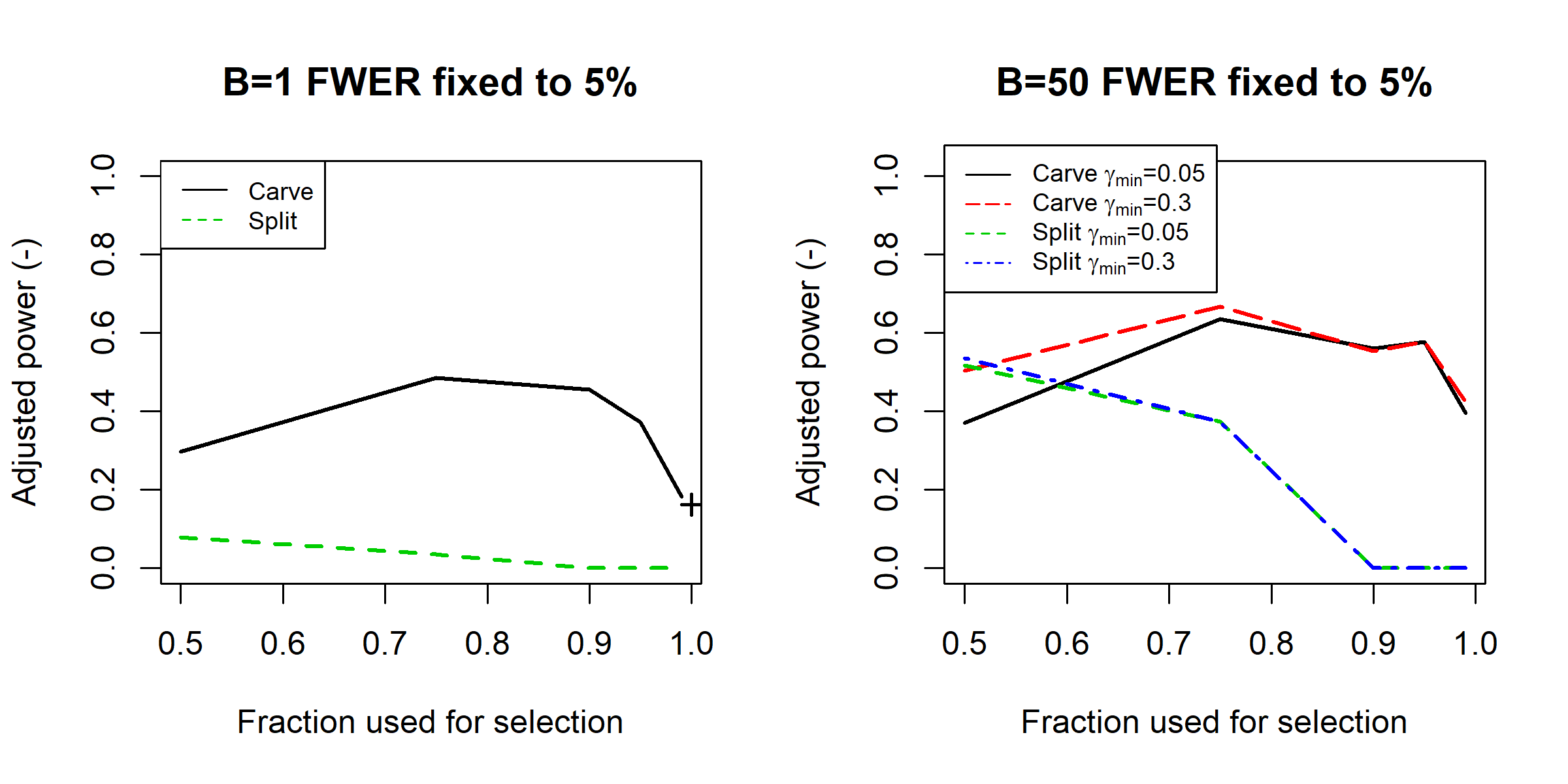}
  \caption[Results for the Toeplitz $X$ in logistic regression.: adjusted results]
  {Results for the Toeplitz design in logistic regression. See caption of Figure \ref{fig:toepadj}.}
  \label{fig:binadj}

\end{figure}
We conduct a similar simulation study as in Section \ref{res:mdc} for the logistic model \eqref{eq:logmod}. We reuse the matrix $X$ coming from a Toeplitz covariance design from Section \ref{res:mdc} with dimensions $n=100$ and $p=200$. The active variables are $\left\{1,5,10,15,20\right\}$, each of which having a coefficient of $2$. After having noticed that cross-validated Lasso tends to select overly sparse models in logistic regression, at least in this set-up, we alter the selection technique. Namely, we select a Lasso model with a given number of selected variables or if there is no such model, the largest model with fewer variables. Inspired by \cite{meinshausen2009p}, we choose this number to be $\left \lfloor{\tfrac{n}{6}}\right \rfloor=16$. Just as for cross-validated Lasso, this introduces a slight bias to our test as $\lambda$ is determined in a data-dependent fashion and is not predefined. We stick to our usual tuning parameters, i.e.,\ $B$ is varied in $\left\{1,10,20,30,40,50\right\}$ and $f$ in $\left\{0.5,0.75,0.9,0.95,0.99,1\right\}$. The target level for the FWER remains at $\alpha=5\%$.

Figures \ref{fig:binunadj} (FWER and power) and \ref{fig:binadj} (adjusted power) illustrate the same performance statistics as for the simulation examples in Section \ref{res:mdc}. Every performance measure corresponds to $200$ simulation runs.

All methods are rather conservative in this set-up. Especially, no value of the FWER above the $5\%$ level occurs. Furthermore, no significant findings are observed for splitting which results in a power of $0$. There exist probably better algorithms for calculating low-dimensional p-values in logistic regression than the ones used for splitting here. Though, as it is not of primary interest to our work, we did not investigate this further. Single-carving has clearly higher power than multicarving, whereas the latter controls the error on a more conservative level. The highest power obtained is $0.28$ (single-carving), $0.16$ ($\gamma_{min}=0.3$) and $0.14$ ($\gamma_{min}=0.05$). All these maxima are reached at $f=0.75$. Pure post-selection inference has a power of $0.088$. Thus, the conjecture that the constraints might be too restrictive is confirmed.

For the trade-off between power and error control, we consider the adjusted power as defined in Section \ref{res:mdc}. Interestingly, multisplitting is now quite competitive. The interpretation is that although p-values are generally larger than $5\%$, there is still a distinction between active and non-active variables. The best adjusted power of the multisplit method is $0.54$. As the curve seems to increase towards lower values of $f$, we further tested $f=0.3$ and $f=0.4$. Neither leads to an increase in the adjusted power for multisplitting such that we can assume that the optimum is reached around $f=0.5$. Multicarving clearly outperforms single-carving with the respective maxima being at $0.67$ ($\gamma_{min}=0.3$), $0.64$ ($\gamma_{min}=0.05$) and $0.49$ (single-carving). Pure post-selection obtains an adjusted power of $0.16$.

In summary, we can state for this data that either of the carving methods improves on pure post-selection inference. The choice between multicarving and single-carving is a trade-off between power and FWER. Our definition of adjusted power, which makes the different methods have equal FWER, is in favor of multicarving.

\subsection{Runtime considerations}\label{res:runtime}
Our method is computationally quite involved while performing empirically well. 
Details are discussed in the Appendix \ref{app:runtime}. The computational 
bottleneck is the MCMC sampling required to calculate p-values and therefore, 
we ignore the other steps for our considerations.
An approximate bound is $\mathcal{O}\left(B\EE\left[{\tilde{s}^4}\right]\right)$ 
for multicarving, where the expectation is due to the fact that $\tilde{s}$ is 
non-constant over splits.

Another popular inference technique for high-dimensional statistics is the de-biased Lasso \citep{van2014asymptotically}. A total of $p+1$ Lasso fits have to be calculated on the entire data. Thus, it scales as $\mathcal{O}\left(p^2\right)$. For high-dimensional data with $p\gg n$ and the standard assumption $\tilde{s}\leq n_1 \leq n$ on the Lasso, we have accordingly $\tilde{s}\ll p$. Then, our multicarve method is more efficient than the de-biased Lasso for $p\rightarrow \infty$ if $n = { \scriptstyle \mathcal{O}}\big(p^{1/2}\big)$.

\section{Discussion and conclusions}
We provide new developments based on the idea of data carving
\citep{fithian2014optimal}. Particularly for high-dimensional scenarios, we
improve upon standard data carving.

First, we introduce multicarving in
the spirit of multisplitting. Our simulation study shows that multicarving generally
leads to better error control and its adjusted power is better than for the single-carve method. Furthermore,
multisplitting and multicarving not only aim to reduce the FWER but also to
make results more replicable. It is very plausible that our multicarve method clearly increases 
replicability compared to single-carving, due to the instability of the Lasso
model selector.

Second, we present group inference, a natural extension of single variable
testing. Such a group test can be applied using single-carving or using the
advocated multicarving. In simulation examples, either variant appears to be
competitive to several methods discussed in \cite{guo2019group}.

Last, we adapt data carving to make it applicable to logistic
linear regression and other generalized linear models. Those adjustments
are based on the central limit theorem and follow from similar 
ideas as already introduced for low-dimensional data and for pure
post-selection inference. Our simulation study leads to the same
conclusions as for the linear model. In particular, data
(multi)carving in the logistic case leads as well to a performance increase compared to pure post-selection
inference.

User-friendly \textsf{R}-software for all of the described (multi)carving methods is available on GitHub, see \textbf{\href{https://github.com/cschultheiss/Multicarving}{https://github.com/cschultheiss/Multicarving}}.

\section*{Acknowledgment}
The research of P. B\"uhlmann was supported in part by the European Research Council under the Grant Agreement No 786461 (CausalStats - ERC-2017-ADG). 

\bibliographystyle{apalike} 
\bibliography{references}

\clearpage

\addtocontents{toc}{\vspace{.5\baselineskip}}
\appendix
\section{Proofs} \label{app:proofs}
\paragraph{Proof of Lemma \ref{lemm:group}}
We require Assumption \eqref{ass:tild2} such that $\big(X_{1,\tilde{S}}^{\top}X_{1,\tilde{S}}\big)^{-1}$ is defined. As in Section \ref{prop:mdc}, we implicitly assume $\text{rank}\big(X_{1,\tilde{S}}\big)=\tilde{s}$ to follow from the sparsity condition. This inverse is implicitly included in $A$ and $\mathbf{b}$. Using the screening assumption, we know $\EE\left[\mathbf{Y}\right]=X_{\tilde{S}}\boldsymbol{\beta}^{\tilde{S}}$. Thus, we can write the unconditional distribution of $\mathbf{Y}$ as follows
\begin{equation*}
\mathbf{Y} \sim \text{exp}\left\{\dfrac{1}{\sigma^2}\left(X_{\tilde{S}}\boldsymbol{\beta}^{\tilde{S}}\right)^{\top}\mathbf{y}-\dfrac{1}{2\sigma^2}\left\Vert\mathbf{y}\right\Vert^2-c\left(X_{\tilde{S}}\boldsymbol{\beta}^{\tilde{S}},\sigma^2\right)\right\},
\end{equation*}
where $c\left(X_{\tilde{S}}\boldsymbol{\beta}^{\tilde{S}},\sigma^2\right)$ denotes the normalizing constant of the Gaussian distribution. We see that $X_{\tilde{S}}^\top \mathbf{Y}$ is the sufficient statistic, while $\boldsymbol{\beta}^{\tilde{S}}$ is the natural parameter as $\sigma$ is assumed to be known. Conditioning on the selection event $\mathbf{Y}\ \big| A\mathbf{Y}\leq\mathbf{b}$ leads to a different exponential family with the same sufficient statistic $X_{\tilde{S}}^\top \mathbf{Y}$ and natural parameter $\boldsymbol{\beta}^{\tilde{S}}$ but different normalizing constant, say, $c^{'}$, compare with \citet[Section 3]{fithian2014optimal}.
\begin{align*}
\mathbf{Y}\ \big| A\mathbf{Y}\leq\mathbf{b}& \sim \text{exp}\left\{\dfrac{1}{\sigma^2}\left(X_{\tilde{S}}\boldsymbol{\beta}^{\tilde{S}}\right)^{\top}\mathbf{y}-\dfrac{1}{2\sigma^2}\left\Vert\mathbf{y}\right\Vert^2-c^{'} \left(X_{\tilde{S}}\boldsymbol{\beta}^{\tilde{S}},\sigma^2\right)\right\}\mathlarger{\mathlarger{\mathbbm{1}}}_{A\mathbf{y}\leq\mathbf{b}}\\
& =\text{exp}\left\{\dfrac{1}{\sigma^2}\left(X_{\tilde{G}}{\boldsymbol{\beta}}_{\tilde{G}}^{\tilde{S}}\right)^{\top}\mathbf{y}+\dfrac{1}{\sigma^2}\left(X_{\tilde{S}\setminus \tilde{G}}\boldsymbol{\beta}^{\tilde{S}}_{-\tilde{G}}\right)^{\top}\mathbf{y}-\dfrac{1}{2\sigma^2}\left\Vert\mathbf{y}\right\Vert^2-c^{'} \left(X_{\tilde{S}}\boldsymbol{\beta}^{\tilde{S}},\sigma^2\right)\right\}\mathlarger{\mathlarger{\mathbbm{1}}}_{A\mathbf{y}\leq\mathbf{b}}\\
& =\text{exp}\left\{\dfrac{1}{\sigma^2}\big({\boldsymbol{\beta}}_{\tilde{G}}^{\tilde{S}}\big)^{\top}X_{\tilde{G}}^{\top}\mathbf{y}+\dfrac{1}{\sigma^2}\big(\boldsymbol{\beta}^{\tilde{S}}_{-\tilde{G}}\big)^{\top}\big(X_{\tilde{S}\setminus \tilde{G}}\big)^{\top}\mathbf{y}-\dfrac{1}{2\sigma^2}\left\Vert\mathbf{y}\right\Vert^2-c^{'} \left(X_{\tilde{S}}\boldsymbol{\beta}^{\tilde{S}},\sigma^2\right)\right\}\mathlarger{\mathlarger{\mathbbm{1}}}_{A\mathbf{y}\leq\mathbf{b}}.
\end{align*}
Here, we split into the parameter that we want to perform inference for ${\boldsymbol{\beta}}_{\tilde{G}}^{\tilde{S}}$ and the nuisance parameter in the model ${\boldsymbol{\beta}}_{-\tilde{G}}^{\tilde{S}}$. From the theory of exponential families, we know that the conditional law $X_{\tilde{G}}^{\top}\mathbf{Y}\ \Big|\Big( \big(X_{\tilde{S}\setminus \tilde{G}}\big)^{\top}\mathbf{Y},A\mathbf{Y}\leq\mathbf{b}\Big)$ does not depend on ${\boldsymbol{\beta}}_{-\tilde{G}}^{\tilde{S}}$. We now want to establish the same result for $\big(X_{\tilde{S}}^+\big)_{\tilde{G}}\mathbf{Y}\ \Big|\Big( \big(X_{\tilde{S}\setminus \tilde{G}}\big)^{\top}\mathbf{Y},A\mathbf{Y}\leq\mathbf{b}\Big)$. For simplicity, we assume $X_{\tilde{S}}=\begin{pmatrix}
X_{\tilde{S} \setminus \tilde{G}}& X_{\tilde{G}}
\end{pmatrix}$ such that it can be separated into variables being part of the group and the others. The result holds w.l.o.g., since permutations of the matrix' columns do not change our inference statement. Then, we get
\begin{align*}
X_{\tilde{S}}^+\mathbf{Y}& =\left(X_{\tilde{S}}^{\top}X_{\tilde{S}}\right)^{-1}X_{\tilde{S}}^{\top}\mathbf{Y}\\
& =\left(X_{\tilde{S}}^{\top}X_{\tilde{S}}\right)^{-1}\begin{pmatrix}
X_{\tilde{S} \setminus \tilde{G}} & X_{\tilde{G}}
\end{pmatrix}^{\top}\mathbf{Y}\\
& =\left(X_{\tilde{S}}^{\top}X_{\tilde{S}}\right)^{-1}\begin{pmatrix}
\big(X_{\tilde{S} \setminus \tilde{G}}\big)^{\top} \\ \big( X_{\tilde{G}}\big)^{\top}
\end{pmatrix}\mathbf{Y}\\
& =\left(X_{\tilde{S}}^{\top}X_{\tilde{S}}\right)^{-1}\begin{pmatrix}
\big(X_{\tilde{S} \setminus \tilde{G}}\big)^{\top}\mathbf{Y} \\ \big( X_{\tilde{G}}\big)^{\top}\mathbf{Y}
\end{pmatrix}.
\end{align*}
Thus, $\big(X_{\tilde{S}}^+\big)_{\tilde{G}}\mathbf{Y}\ \Big|\Big( \big(X_{\tilde{S}\setminus \tilde{G}}\big)^{\top}\mathbf{Y},A\mathbf{Y}\leq\mathbf{b}\Big)$ is a fixed affine transform of\\
$X_{\tilde{G}}^{\top}\mathbf{Y}\ \Big|\Big( \big(X_{\tilde{S}\setminus \tilde{G}}\big)^{\top}\mathbf{Y},A\mathbf{Y}\leq\mathbf{b}\Big)$, making it independent from ${\boldsymbol{\beta}}_{-\tilde{G}}^{\tilde{S}}$ as well. Naturally, the subset $\big(X_{\tilde{S}}^+\big)_{\tilde{G}}\mathbf{Y}$ is conditionally independent too. Based on our two assumptions, the only parameters in the model are ${\boldsymbol{\beta}}_{-\tilde{G}}^{\tilde{S}}$ and ${\boldsymbol{\beta}}_{\tilde{G}}^{\tilde{S}}$. Thus, after establishing independence from the former, the only parameter left in the model is the latter, which is exactly the lemma's statement.
\paragraph{Proof of Theorem \ref{theo:grouperror}}
For a group $G$, we either have $\big\vert\tilde{G}\big\vert>0$ or $\big\vert\tilde{G}\big\vert=0$. Assume the former case first. Due to screening, we know $\beta^{\tilde{S}}_j=\beta_j \ \forall j\in\tilde{S}$, which leads to $\beta^{\tilde{S}}_j=\beta_j \ \forall j\in\tilde{G}$ as $\tilde{G}\subseteq \tilde{S}$. Null hypothesis \eqref{eq:fullnullgroup} then directly implies
\begin{equation*}
\beta_j=0 \ \forall j \in G \, \rightarrow \, \beta_j=0 \ \forall j \in \tilde{G} \, \rightarrow \, \beta_j^{\tilde{S}}=0 \ \forall j \in \tilde{G},
\end{equation*}
which corresponds to null hypothesis \eqref{eq:selectednullgroup}. Therefore, all assumptions of Theorem \ref{theo:groupunif} are fulfilled, leading to the uniform distribution of the p-value. Error control can thus be stated as
\begin{equation*}
\PR{p_{G}\left(Y\right)\leq \alpha}=\PR{p_{\tilde{G}}\left(Y\right)\leq \alpha}=\alpha\leq \alpha.
\end{equation*}
In the other case ($\big\vert\tilde{G}\big\vert=0$) we have
\begin{equation*}
\PR{p_{G}\left(Y\right)\leq \alpha}=0\leq \alpha.
\end{equation*}
Thus, we obtain error control in either case, which closes the proof.
\section{Sampling from a linearly constrained Gaussian}
\label{app:mcmc}
The algorithm presented in this section is strongly based on the GitHub repository cited in \cite{fithian2014optimal} for their simulations. However, since there seems to be no written documentation of the algorithm itself and the theory behind, we provide it for the interested reader.

For simplicity, we will suppress index $\tilde{S}$, since we implicitly assume to work in a selected submodel throughout this section.

In order to do inference for variable $j$, the goal is to sample from $\mathbf{Y}\sim\mathcal{N}\left(X\boldsymbol{\beta},\sigma^2 I_n\right)$ subject to $A\mathbf{Y}\leq\mathbf{b}$, $\left(X_{-j}\right)^{\top}\mathbf{Y}=\left(X_{-j}\right)^{\top}\mathbf{y} \equiv\mathbf{d}$ and $\beta_j=0$.
The first condition leads to boundaries on the sampling region, the second one changes both the mean parameter and the covariance matrix, and the last one further changes the mean and creates a null distribution.
\subsection{Change of mean and covariance}
\label{app:eqcon}
Let $\mathbf{Z}$ be a Gaussian random vector with mean $\boldsymbol{\mu}$ and covariance $\Sigma$. We are interested in $\mathbb{E}\left[\mathbf{Z}\ \Big| C\mathbf{Z}=\mathbf{d}\right] \equiv \tilde{\boldsymbol{\mu}}$ and $\text{Cov}\left(\mathbf{Z}\ \Big| C\mathbf{Z}=\mathbf{d}\right)\equiv \tilde{\Sigma}$. To find those, split $\mathbf{Z}$ into
\begin{equation*}
\mathbf{Z}=\Sigma C^{\top}\left( C\Sigma C^{\top}\right)^{-1} C\mathbf{Z}+\left( I-\Sigma C^{\top}\left( C\Sigma C^{\top}\right)^{-1} C\right)\mathbf{Z}.
\end{equation*}
One can see (e.g.,\ by calculating the covariance) that the second term is independent of $ C\mathbf{Z}$, thus unchanged by the conditioning, while the first part is completely defined by the conditioning. Thus, we have 
\begin{align*}
\tilde{\boldsymbol{\mu}}& =\mathbb{E}\left[\mathbf{Z}\ \Big| C\mathbf{Z}=\mathbf{d}\right]\\
& =\mathbb{E}\left[\Sigma C^{\top}\left( C\Sigma C^{\top}\right)^{-1} C\mathbf{Z}+\left( I-\Sigma C^{\top}\left( C\Sigma C^{\top}\right)^{-1} C\right)\mathbf{Z}\ \Big| C\mathbf{Z}=\mathbf{d}\right]\\
& =\Sigma C^{\top}\left( C\Sigma C^{\top}\right)^{-1}\mathbf{d}+\mathbb{E}\left[\left( I-\Sigma C^{\top}\left( C\Sigma C^{\top}\right)^{-1} C\right)\mathbf{Z}\right]\\
& =\Sigma C^{\top}\left( C\Sigma C^{\top}\right)^{-1}\mathbf{d}+\left( I-\Sigma C^{\top}\left( C\Sigma C^{\top}\right)^{-1} C\right)\boldsymbol{\mu}
\end{align*}
and similarly
\begin{align*}
\tilde{\Sigma}& = \text{Cov}\left(\mathbf{Z}\ \Big| C\mathbf{Z}=\mathbf{d}\right)\\
& = \text{Cov}\left(\Sigma C^{\top}\left( C\Sigma C^{\top}\right)^{-1} C\mathbf{Z}+\left( I-\Sigma C^{\top}\left( C\Sigma C^{\top}\right)^{-1} C\right)\mathbf{Z}\ \Big| C\mathbf{Z}=\mathbf{d}\right)\\
& = \text{Cov}\left(\Sigma C^{\top}\left( C\Sigma C^{\top}\right)^{-1} C\mathbf{Z}\ \Big| C\mathbf{Z}=\mathbf{d}\right)+\text{Cov}\left(\left(I-\Sigma C^{\top}\left( C\Sigma C^{\top}\right)^{-1} C\right)\mathbf{Z}\ \Big| C\mathbf{Z}=\mathbf{d}\right)+\\
& \quad \ 2\text{Cov}\left(\Sigma C^{\top}\left( C\Sigma C^{\top}\right)^{-1} C\mathbf{Z},\ \left(I-\Sigma C^{\top}\left( C\Sigma C^{\top}\right)^{-1} C\right)\mathbf{Z}\ \Big| C\mathbf{Z}=\mathbf{d}\right)\\
& = 0+\text{Cov}\left(\left( I-\Sigma C^{\top}\left( C\Sigma C^{\top}\right)^{-1} C\right) \mathbf{Z}\right)+0\\
& = \Sigma-\Sigma C^{\top}\left( C\Sigma C^{\top}\right)^{-1} C\Sigma.
\end{align*}

In our problem of interest, we have $\boldsymbol{\mu}=X_{-j}\boldsymbol{\beta}_{-j}$ (after setting $\beta_j=0$), $\Sigma=\sigma^2 I_n$, and $ C=\left(X_{-j}\right)^{\top}$. This yields
\begin{equation*}
\tilde{\boldsymbol{\mu}}=X_{-j}\left(X_{-j}^{\top}X_{-j}\right)^{-1}\mathbf{d}=P_{X_{-j}}\mathbf{y}
\end{equation*}
and
\begin{equation*}
\tilde{\Sigma}=\sigma^2\left( I_n-X_{-j}\left(X_{-j}^{\top}X_{-j}\right)X_{-j}^{\top}\right)=\sigma^2 P_{X_{-j}}^{\perp}.
\end{equation*}
Most importantly, the mean term does not have any dependence on $\boldsymbol{\beta}_{-j}$ such that we can calculate an inference statement without knowing the other coefficients.

\subsection{Computational shortcuts: linear transformations} \label{app:lintrafo}
Since all constraints are linear, they can also be guaranteed for linear transformations of $\mathbf{Y}$ if not too much dimensionality reduction is applied.

Define the least squares solution on all data as 
\begin{equation*}
\widehat{\boldsymbol{\beta}}=\left(X^{\top}X\right)^{-1}X^{\top}\mathbf{Y}
\end{equation*}
and the one on the selection data only as 
\begin{equation*}
\widehat{\boldsymbol{\beta}}_1=\left(X_1^{\top}X_1\right)^{-1}X_1^{\top}\mathbf{Y}_1.
\end{equation*}
Then, two vectors which are well suited to fulfil all constraints after transformation are
\begin{equation*}
\mathbf{U}=\begin{pmatrix} \widehat{\boldsymbol{\beta}}\\ \widehat{\boldsymbol{\beta}}_1 \end{pmatrix} \in \mathbb{R}^{2\tilde{s}} \quad \text{or} \quad \mathbf{V}=\begin{pmatrix} \widehat{\boldsymbol{\beta}}_1 \\\mathbf{Y}_2 \end{pmatrix}\in \mathbb{R}^{\tilde{s}+n_2}.
\end{equation*}
Since those are linear transformations, they will still be Gaussian with mean and covariance that can be easily derived from those of $\mathbf{Y}$.

Further, the constraints transform to
\begin{equation*}
X_{-j}^{\top}\mathbf{Y}=\mathbf{d} \, \leftrightarrow \, \begin{pmatrix}
X^{\top}X & {0}_{\tilde{s}\times\tilde{s}}
\end{pmatrix}_{\left[\left\{1,\ldots,\tilde{s}\right\}\setminus j,\ \left\{1,\ldots,2\tilde{s}\right\}\right]}\mathbf{U}=\mathbf{d}
\end{equation*}
\begin{equation*}
X_{-j}^{\top}\mathbf{Y}=\mathbf{d} \, \leftrightarrow \, \begin{pmatrix}
X_1^{\top}X_1 & X_2^{\top}
\end{pmatrix}_{\left[\left\{1,\ldots,\tilde{s}\right\}\setminus j,\ \left\{1,\ldots,\left(\tilde{s}+n_2\right)\right\}\right]}\mathbf{V}=\mathbf{d}.
\end{equation*}
We use the bracket notation for the indices to indicate that row $j$ of the resulting matrix has to be omitted.
And, by using the active constraints from \cite{lee2016exact}, we have 
\begin{equation*}
A=-\text{diag}\left(\widehat{\boldsymbol{\xi}}\right)\left(X_1^{\top}X_1\right)^{-1}X_1^{\top} \quad , \quad \mathbf{b}=- \lambda \text{diag}\left(\widehat{\boldsymbol{\xi}}\right)\left(X_1^{\top}X_1\right)^{-1}\widehat{\boldsymbol{\xi}},
\end{equation*}
where $\widehat{\boldsymbol{\xi}}$ denotes the signs of the parameters' Lasso estimates. This can be transformed to
\begin{equation*}
A\mathbf{Y}_1\leq\mathbf{b}\leftrightarrow \begin{pmatrix}
0_{\tilde{s}\times \tilde{s}} && -\text{diag}\left(\widehat{\boldsymbol{\xi}}\right)
\end{pmatrix}\mathbf{U}\leq\mathbf{b}
\end{equation*}
\begin{equation*}
A\mathbf{Y}_1\leq\mathbf{b}\leftrightarrow \begin{pmatrix}
-\text{diag}\left(\widehat{\boldsymbol{\xi}}\right) && 0_{\tilde{s}\times \tilde{n_2}}
\end{pmatrix}\mathbf{V}\leq\mathbf{b}.
\end{equation*}
Thus, we have transformed the linear equality and inequality constraints and can proceed as if we were to sample from $\mathbf{Y}$ by firstly adjusting the mean and the covariance matrix as described in Section \ref{app:eqcon}.

The choice of whether to sample from $\mathbf{U}$ or $\mathbf{V}$ is rather simple: just use whichever has lower dimensionality in order to increase efficiency. As stated in Section \ref{datacarving}, one would further condition on $\left\Vert\mathbf{Y}\right\Vert^2$ in the unknown variance case. Though, this constraint is not transformable to $\mathbf{U}$ or $\mathbf{V}$, thus the dimensionality could not be reduced. Therefore, we use an estimate of the variance instead of the (theoretically beautiful) conditioning idea for our simulations.

\subsection{Whitening}\label{app:white}
In order to make the MCMC algorithm simpler, we would like to always sample from zero mean unit variance independent Gaussians (i.e.,\ white Gaussians). This can be achieved by applying a further linear transformation. We need a forward map transforming the initial point and an inverse map transforming back the MCMC sample.

Assume that we sample from $\mathbf{Y}\sim \mathcal{N}\left(\boldsymbol{\mu},\Sigma\right)\ \big| A\mathbf{Y}\leq\mathbf{b}$ which is achieved by applying the transformations from the previous two sections. Here, $\Sigma \in \mathbb{R}^{n\times n}$ has rank $r=n+1-\tilde{s}$, i.e.,\ $\Sigma$ is not full-ranked whenever $\tilde{s}>1$. This is as we lose some degrees of freedom after conditioning (cf.\ Section \ref{app:eqcon}). Further, define matrices $\Sigma^{\frac{1}{2}}\in\mathbb{R}^{n\times r}$ and $\Sigma^{-\frac{1}{2}}\in\mathbb{R}^{r\times n}$ such that
\begin{equation*}
\Sigma^{\frac{1}{2}}\left(\Sigma^{\frac{1}{2}}\right)^\top=\Sigma, \quad \Sigma^{-\frac{1}{2}}\Sigma^{\frac{1}{2}}=I.
\end{equation*}
These can be found, e.g.,\ by using the eigenvalue decomposition of $\Sigma$. Then, our forward map is
\begin{equation*}
\mathbf{Y}^{'}=W\big(\mathbf{Y}\big)=\Sigma^{-\frac{1}{2}}\left(\mathbf{Y}-\boldsymbol{\mu}\right),
\end{equation*}
and accordingly, the inverse map is 
\begin{equation*}
\mathbf{Y}=W^{-1}\big(\mathbf{Y}^{'}\big)=\Sigma^{\frac{1}{2}}\mathbf{Y}^{'}+\boldsymbol{\mu}.
\end{equation*}
Note that $W\big(W^{-1}\big(\mathbf{Y}^{'}\big)\big)=\mathbf{Y}^{'} \ \forall \mathbf{Y}^{'}$ and further $W^{-1}\big(W\big(\mathbf{Y}\big)\big)$ for all $\mathbf{Y}$ fulfilling the equality constraints, thus all $\mathbf{Y}$ we are interested in.

Importantly, the boundary constraint $A\mathbf{Y}\leq\mathbf{b}$ has to be transformed as well. This is possible by
\begin{equation*}
A\mathbf{Y}\leq\mathbf{b} \ \leftrightarrow \ A\left(\mathbf{Y}-\boldsymbol{\mu}\right)\leq\mathbf{b}-A\boldsymbol{\mu} \ \leftrightarrow \ A\Sigma^{\frac{1}{2}}\mathbf{Y}^{'}\leq\mathbf{b}-A\boldsymbol{\mu},
\end{equation*}
which leads to 
\begin{equation*}
A^{'}=A\Sigma^{\frac{1}{2}}\ , \quad \mathbf{b}^{'}=\mathbf{b}-A\boldsymbol{\mu},
\end{equation*}
i.e.,\ the constraints in the whitened space. With these whitened constraints at hand, the only thing left is to sample from a white Gaussian subject to linear inequality constraints.

Notably, since $\Sigma^{-\frac{1}{2}}$ is a wide matrix ($r<n$ unless $\tilde{s}=1$), we transform into a lower-dimensional space. Therefore, the transformation into the withened space leads to a further dimensionality reduction, which makes the sampling more efficient.
\subsection{Sampling from a linearly constrained white Gaussian} \label{app:MCMC}
The MCMC algorithm presented in this section is as well based on the mentioned GitHub repository. Though, we emphasize that any algorithm approximating a white Gaussian with linear inequality constraints could be invoked in this place using the same preprocessing steps (cf.\ Sections \ref{app:eqcon} - \ref{app:white}).

For simplicity, reuse all initial names, thus we want to sample from $\mathbf{Y}\sim \mathcal{N}\left(\mathbf{0},I\right)$ subject to $A\mathbf{Y}\leq\mathbf{b}$ and let $\mathbf{y}_0$ be a point fulfilling the constraints. More precisely, $\mathbf{y}_0$ is the preprocessed version of the observed vector.

The idea is to move in every step $t$ in a given random direction $\boldsymbol{\eta}^t$, while keeping the projections into its orthogonal complement fixed, i.e.,
\begin{equation*}
\mathit{P}^{\perp}_{\boldsymbol{\eta}^t}\mathbf{Y}^t=\mathit{P}^{\perp}_{\boldsymbol{\eta}^t}\mathbf{y}^{t-1}.
\end{equation*}
Or in other words, we want to sample from 
\begin{equation*}
\mathbf{Y}^t\sim \mathcal{N}\left(\mathbf{0},I\right) \quad \text{subject to} \quad A\mathbf{Y}^t\leq\mathbf{b}, \quad \mathit{P}^{\perp}_{\boldsymbol{\eta}^t}\mathbf{Y}^t=\mathit{P}^{\perp}_{\boldsymbol{\eta}^t}\mathbf{y}^{t-1}.
\end{equation*}
This is in exact analogy to the set-up in \cite{lee2016exact} for pure post-selection inference using $\boldsymbol{\eta}^t$ as direction of interest and $\mathbf{y}^{t-1}$ as observation to base the inference on. Thus, the boundary derived for pure post-selection inference can be reused, making $\big(\boldsymbol{\eta}^t\big)^{\top}\mathbf{Y}^t$ a univariate truncated Gaussian with known mean and variance. One can easily sample from this leading to a new point $\mathbf{y}^t$. For every $\mathbf{Y}^t$, this can be repeated for a new random direction $\boldsymbol{\eta}^t$ such that the whole constrained space should be explored. After enough steps, the samples should approximate the null distribution sufficiently well.

An alternative algorithm that could be used for the actual MCMC sampling is the Hamiltonian Monte Carlo algorithm described in \cite{pakman2014exact}. An implementation thereof is available in the \textsf{R}-package \texttt{tmg} \citep{pakman2015tmg}.

\section{Additional numerical results} \label{app:num}
This section contains additional numerical results and details about 
runtime considerations. 

\subsection{Multicarving for the linear model} \label{app:mdc}
We consider slight variations of the simulation set-ups in Section \ref{res:mdc}. Especially, we look at scenarios where the selection stage is rather hard leading to low probability of screening. This can have a negative impact on the performance of the inference methods for multiple reasons. First, without screening the theoretical validity for the error control is not given anymore. Second, selecting less true active predictors leads to less potential for true rejections such that the power drops.
\subsubsection{Toeplitz design with different correlation parameter} \label{app:toep}
As we mention in Section \ref{prop:mdc}, the correlation between predictors has a high impact on the success of screening in the finite data set-up and accordingly, on the performance of our procedure. To analyze this effect, we redo our simulation for the Toeplitz design in Section \ref{res:toep} with different correlation parameter $\rho$. We test the values $\rho=0.3$ and $\rho=0.9$ and otherwise proceed as before. We sample the predictor matrix $X$ once for each value of $\rho$. To make things as comparable as possible, we fix the noise level such that $\tfrac{\widehat{\text{Var}}\left(X\boldsymbol{\beta}\right)}{\sigma^2}=1.71$ as it was in the set-up in Section \ref{res:toep}.

\begin{figure}[b!]
  \centering
  \includegraphics[width=1\textwidth]{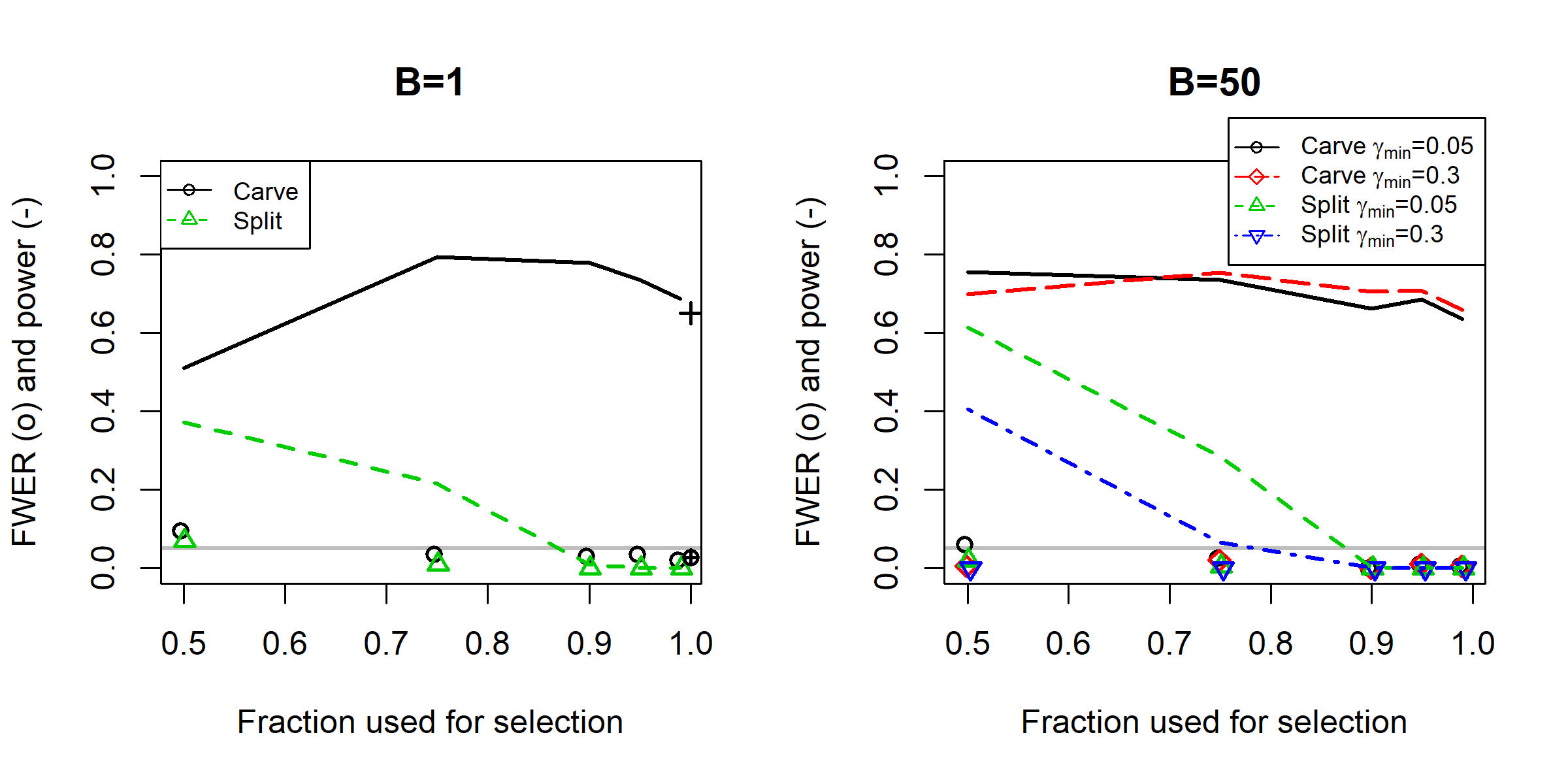}
  \caption[Toeplitz design: unadjusted results]
  {Results for the Toeplitz design with $\rho=0.3$. See caption of Figure \ref{fig:toepunadj}.}
  \label{fig:toep03unadj}
\end{figure}
\begin{figure}[htb]
  \centering
  \includegraphics[width=1\textwidth]{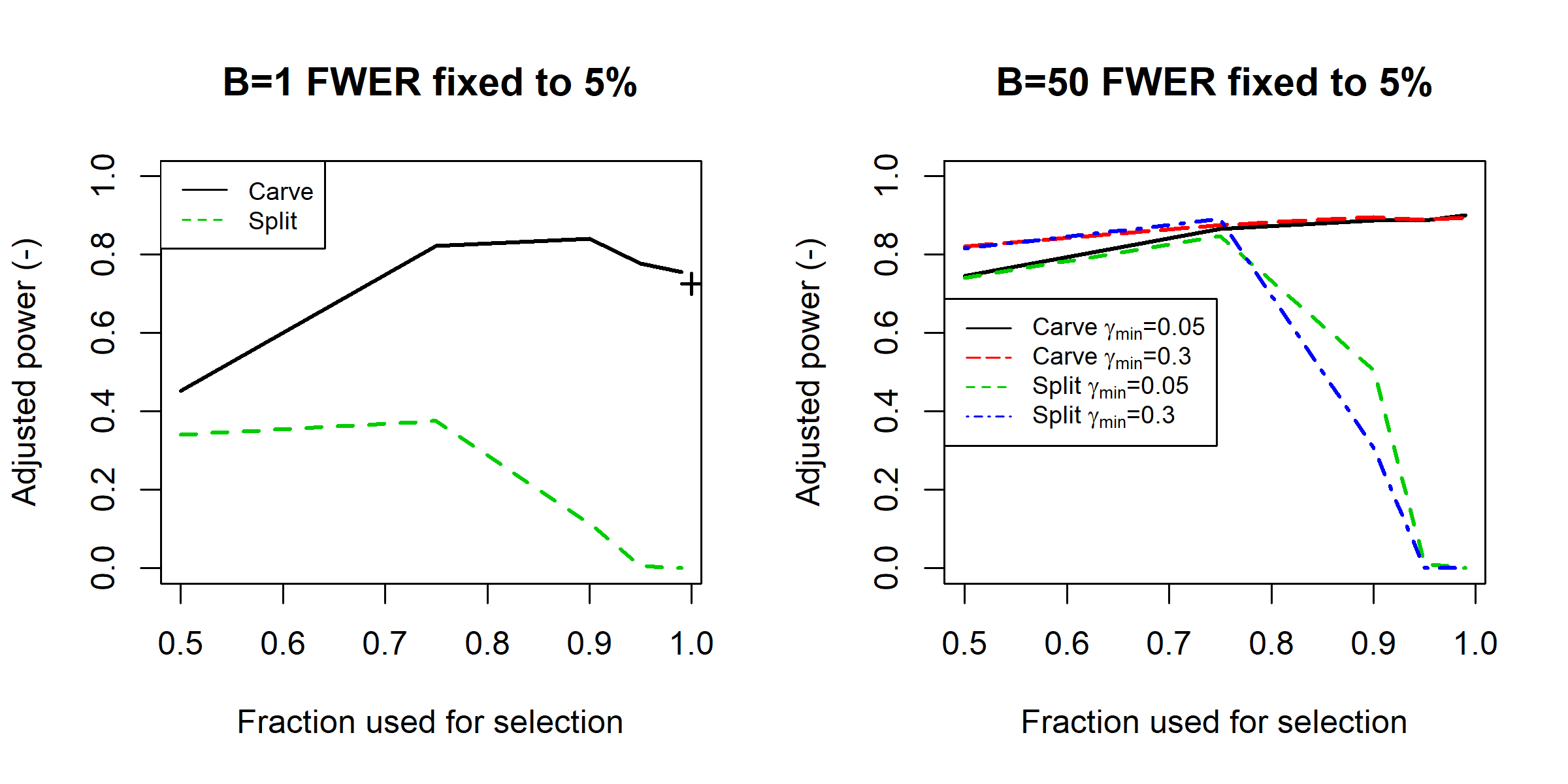}
  \caption[Toeplitz design: adjusted results]
  {Results for the Toeplitz design with $\rho=0.3$ for the adjusted power. See caption of Figure \ref{fig:toepadj}.}
  \label{fig:toep03adj}
\end{figure}
We first consider $\rho = 0.3$ for which we show the obtained FWER and power in Figure \ref{fig:toep03unadj} and the obtained adjusted power in Figure \ref{fig:toep03adj}. Comparing this to our base case in Section \ref{res:toep}, we see that the curves for power and adjusted power are much higher while as the FWER are at a lower level. Thus, this problem is a lot easier to handle by all the inference methods at hand as one would expect. Our conclusions are similar to the set-up with $\rho=0.6$ comparing the different methods. Single-carving obtains the highest power with $\text{FWER} \leq 5 \%$ with a maximum of $0.79$ whereas the multicarving methods reach $0.74$ ($\gamma_{min}=0.05$) and $0.75$ ($\gamma_{min}=0.3$). Though, multicarving controls the error more conservatively leading to better adjusted power with respective maxima of $0.90$ for either carving method and $0.84$ for single-carving. Two things shall be noted: First, $f=0.75$ is now competitive with higher selection fractions which can be explained by the empirical success rate of screening that is already rather high ($76.5 \%$) for $f=0.75$. Second and related, multisplitting is also more competitive since screening works reasonably well for selection fractions for which there is still some power left using only the second part of the data for inference.

\begin{figure}[h]
  \centering
  \includegraphics[width=1\textwidth]{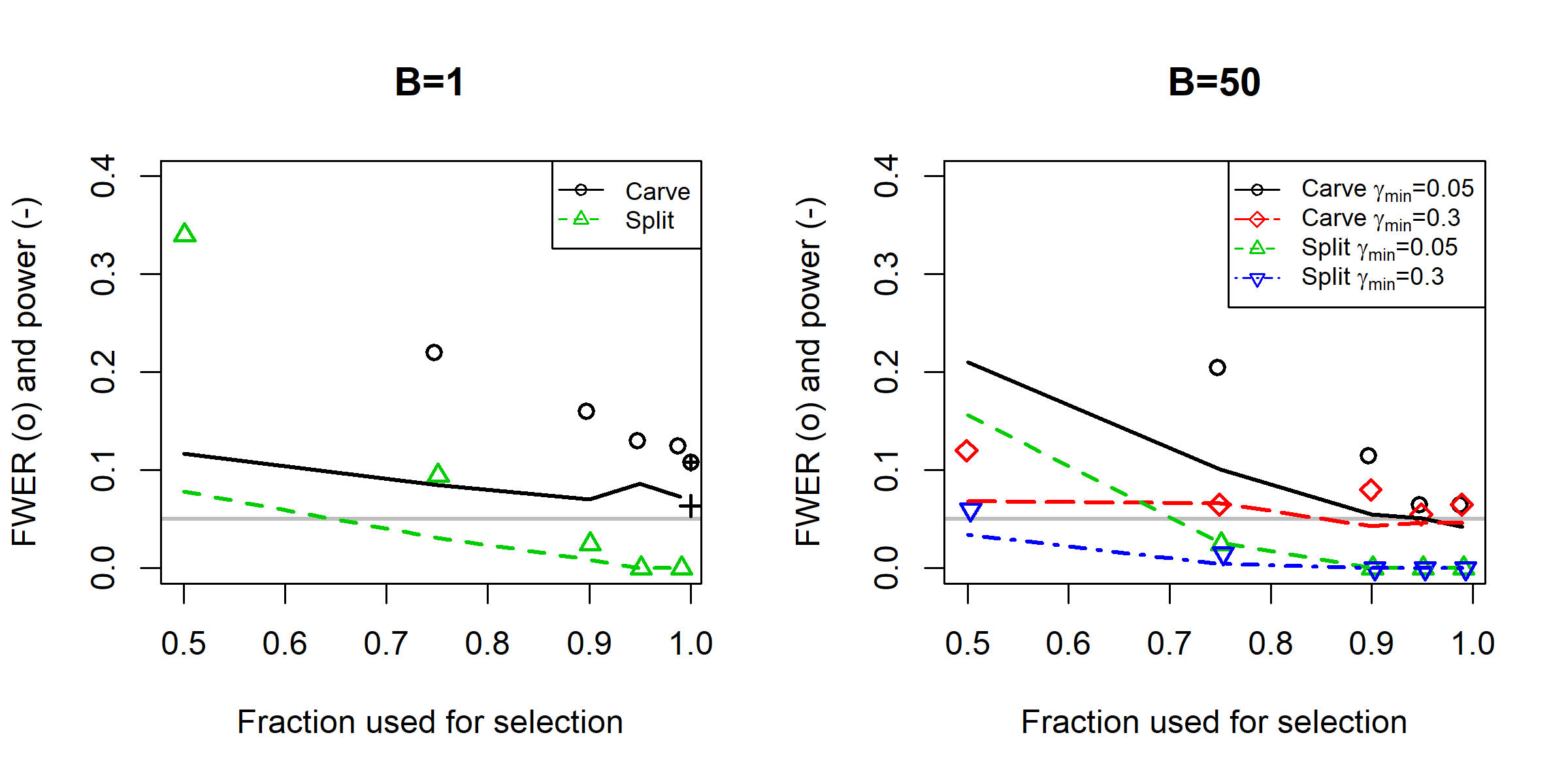}
  \caption[Toeplitz design: unadjusted results]
  {Results for the Toeplitz design with $\rho=0.9$. See caption of Figure \ref{fig:toepunadj}.}
  \label{fig:toep09unadj}
\end{figure}

\begin{figure}[h!]
  \centering
  \includegraphics[width=1\textwidth]{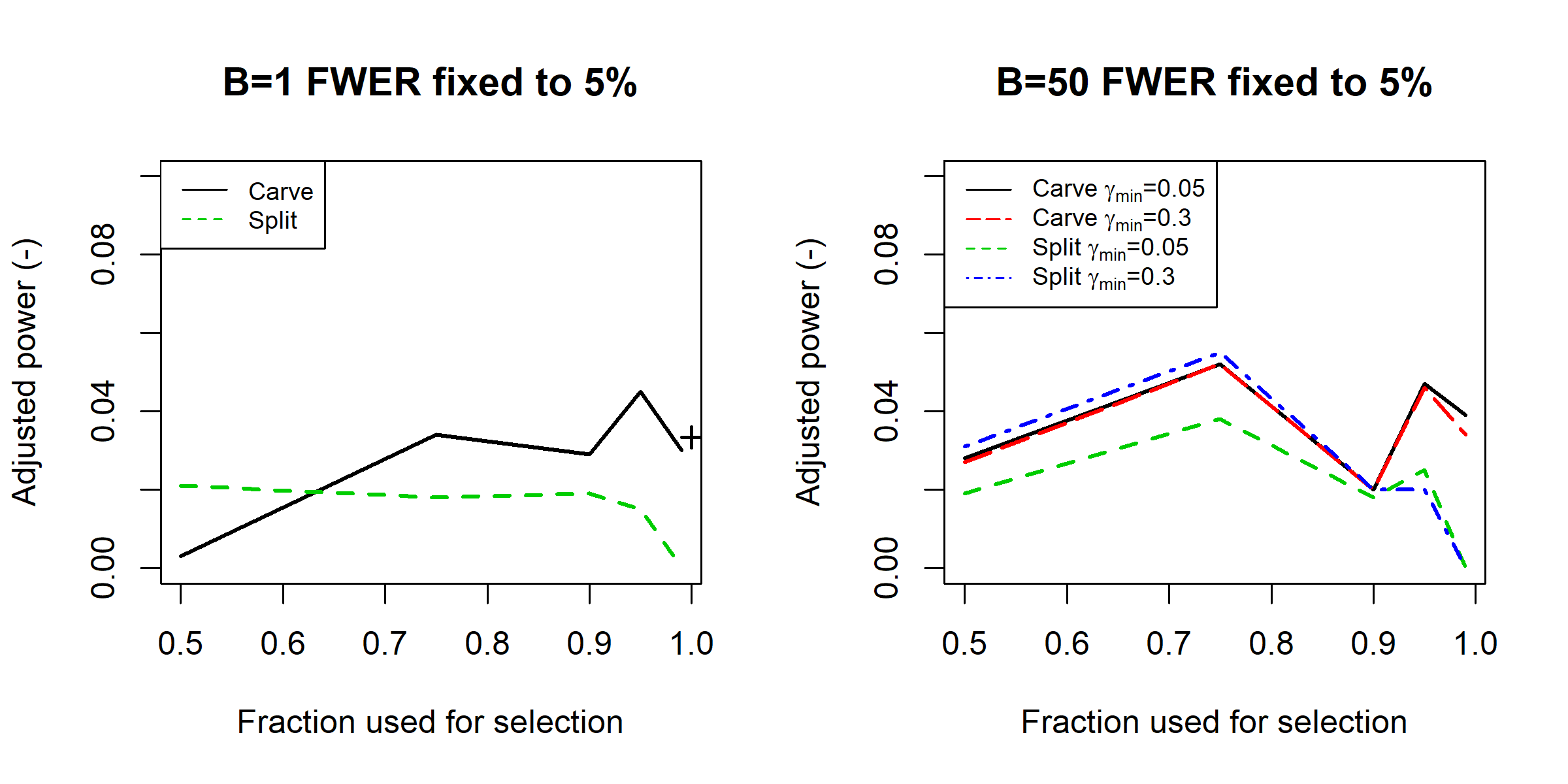}
  \caption[Toeplitz design: adjusted results]
  {Results for the Toeplitz design with $\rho=0.9$ for the adjusted power. See caption of Figure \ref{fig:toepadj}.}
  \label{fig:toep09adj}
\end{figure}

For the high-correlation case with $\rho=0.9$, the results are displayed in Figures \ref{fig:toep09unadj} and \ref{fig:toep09adj}. Note that the plotting range is restricted to a more representative area and that some FWER symbols above the level $40\%$ are thus missing. As expected, those results now look much worse. Especially, neither carving method is able to control the FWER at $5\%$ for any selection fraction. Of course, this relates to the low probability of screening which is only at $7.9\%$ even when using all the data for the selection stage. Nevertheless, we still see that multicarving leads to better error control than single-carving except for $f=0.5$, where the FWER for single-carving is $43\%$ and for $\gamma_{min}=0.05$ it is even $71\%$. Accordingly, the best adjusted power is also better for multicarving. Though, all the values are on a very low level with respective maxima of $0.052$ for either value of $\gamma_{min}$ and $0.045$ for single-carving. Further, we note that multisplitting with $\gamma_{min}=0.3$ performs roughly as well as multicarving with respect to the adjusted power for $f=0.5$ and $f=0.75$. We think that this is because the performance of each method is mainly driven by the selection quality in this scenario such that the blessings of multiplicity are more pronounced than those of carving.

In summary, our assumption that lower correlation leads to better performance and vice-versa is confirmed in this analysis. Especially, none of the inference techniques in scope works well in a scenario where the selection stage is very difficult and screening is very unlikely. Nevertheless, we can still see some positive effect of using multiple splits in this scenario.

\subsubsection{Semi-synthetic Riboflavin data for sparsity 4} \label{app:ribo4}
\begin{figure}[h!]
  \centering
  \includegraphics[width=1\textwidth]{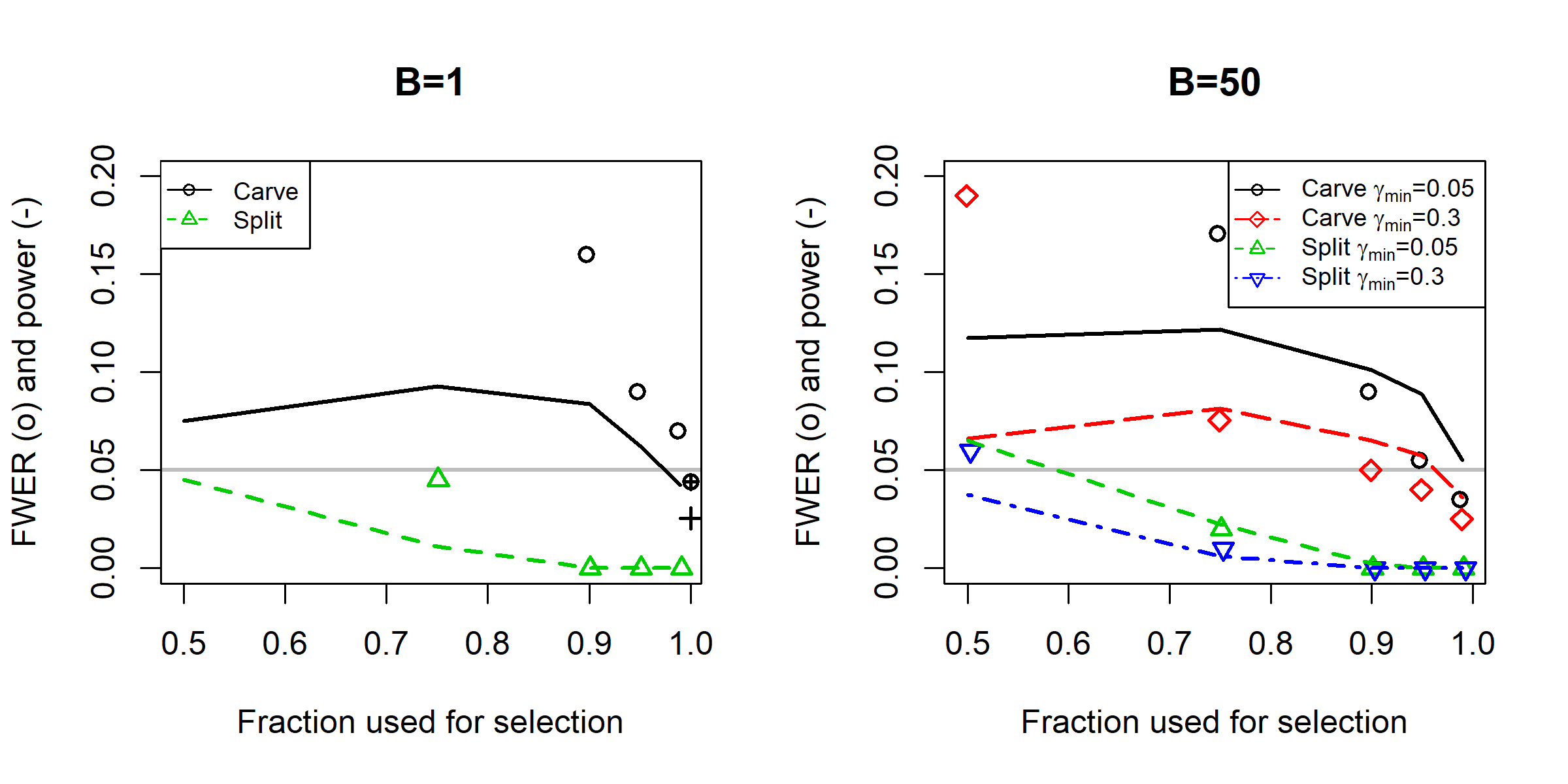}
  \caption[Riboflavin $X$ with sparsity 4: unadjusted results]
  {Results for the Riboflavin $X$ with sparsity 4. See caption of Figure \ref{fig:toepunadj}.}
  \label{fig:ribo4unadj}
    \includegraphics[width=1\textwidth]{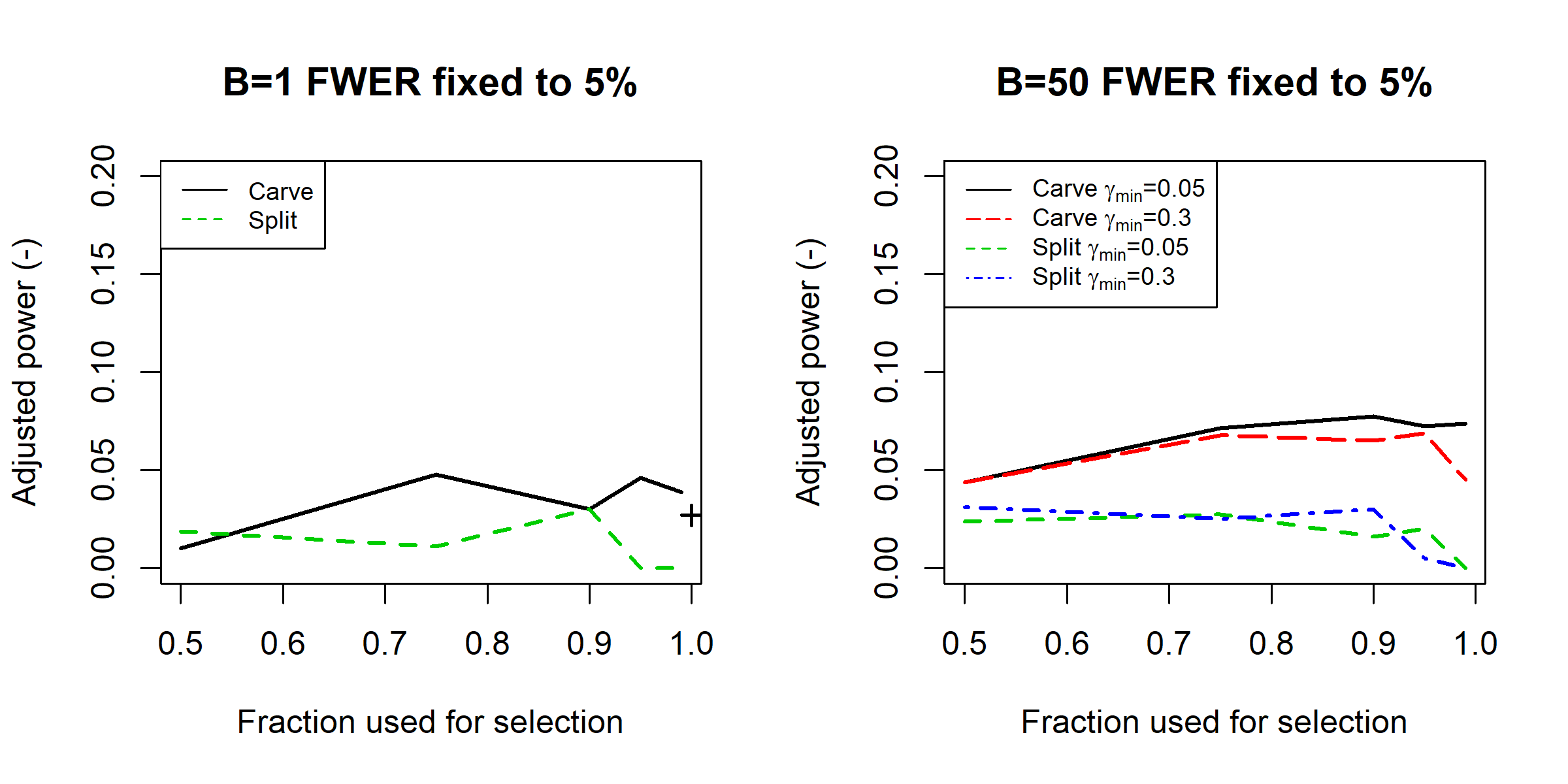}
  \caption[Riboflavin $X$ with sparsity 4: adjusted results]
  {Results for the Riboflavin $X$ with sparsity 4 for the adjusted power. See caption of Figure \ref{fig:toepadj}.}
  \label{fig:ribo4adj}
\end{figure}

We redo the simulation as in Section \ref{res:ribo} setting the sparsity to $4$ without changing anything else. The respective results are presented in Figures \ref{fig:ribo4unadj} (FWER and power) and \ref{fig:ribo4adj} (adjusted power).

Note that we restrict the plotting area of the y-axis to a maximum of $0.2$ such that some values of the FWER are non-visible.
At first glance, one sees that the power is generally quite low for all methods while as the error is above the $5\%$ level for many set-ups leading also to low adjusted power. As in Appendix \ref{app:toep}, this relates to the difficulty for the selection stage. Screening only worked in $9.8\%$ of the simulation runs using all data for selection and naturally even less for any subset. For comparison, screening worked in $81.3\%$ of the instances in the sparser alternative, which makes the problem much easier.

In this set-up, multicarving with $\gamma_{min}=0.05$ has the highest power for all $f$, while $\gamma_{min}=0.3$ has the lowest FWER amongst the three carving methods. The highest power obtained controlling the FWER at $5\%$ is in favor of using $\gamma_{min}=0.3$ with a value of $0.065$. The other two methods obtain respective maxima of $0.055$ ($\gamma_{min}=0.05$) and $0.026$ (single-carving). Especially, single-carving only reaches error control at $f=1$ which is pure post-selection inference. The adjusted power is slightly higher for $\gamma_{min}=0.05$ than for $\gamma_{min}=0.3$ with maximal values of $0.078$ and $0.069$. For single-carving, the maximal value is $0.048$. In summary, multicarving is to be preferred over single-carving in this difficult set-up.

As one of the main difficulties in this scenario is the bad screening property, a natural adaption is the use of $\lambda_{min}$ instead of $\lambda_{1se}$ for selection with cross-validation. This leads to larger selected models and could potentially increase the probability of screening. Our simulation confirms that this leads to a performance boost with the highest adjusted power for multicarving now being $0.148$. For simplicity, we refrain from showing the results in detail. It has to be mentioned though that the use of $\lambda_{min}$ leads to a substantial increase in runtime as more variables are selected. We elaborate this effect further in Section \ref{app:runtime}.

\subsection{Data carving for group testing: sparse scenario}\label{app:group}
We refer to Section \ref{res:group} for more details about the implementation and further discussion. 

For the sparse scenario, we choose $\boldsymbol{\beta}$ to be sparse and the active covariates are strongly correlated with other covariates. The number of covariates $p$ is as well $500$, and $X$ is simulated using the following covariance structure
\begin{equation*}
\Sigma_{jl}=
\begin{cases}
0.8 & \text{if} \ 1\leq j \neq l \leq 5 \\
0.6^{\left\vert j-l\right\vert} & \text{otherwise}.
\end{cases}
\end{equation*}
Thus, $\Sigma$ is the same Toeplitz matrix as in the dense alternative described in Section \ref{res:group} unless for the first five variables. The parameter vector is defined as $\beta_1=\beta_3=\delta$ and $\beta_j=0$ otherwise, meaning that the active variables are within the highly correlated set.
This time $\delta$ is varied over $\left\{0,0.1,0.2,0.3,0.4,0.5\right\}$ and $n$ over $\left\{250,350,500\right\}$. The response $\mathbf{Y}$ is generated as before, leading to SNR in $\{0,0.036,0.144,0.324,$ $0.576,0.9\}$. In this scenario, we are interested in the null hypothesis \eqref{eq:fullnullgroup} for the group $G=\left\{1,2, \ldots ,5\right\}$.

\subsubsection{Single-carving for group testing: sparse scenario}
In Table \ref{tab:sparse}, we report the empirical rejection rate for the scenario with very sparse $\boldsymbol{\beta}$ and highly correlated features.
This scenario seems to be easier to handle than the dense
scenario. Especially, the error is controlled at a more conservative level,
with the highest error being $1.5\%$. For the power, the tendencies are
similar as before. For $\delta \in \left[0.1,0.2\right]$, $f=0.5$ and $f=1$
have generally the lowest power, while the highest power is obtained with
$f \in \left[0.75,0.95\right]$. Starting from $\delta=0.3$, $f=1$ leads to
the lowest power, while the other ERR are mostly exactly $1$.
\begin{table}[!htb]
\centering
\begin{tabular}{|lr|cccccc|}
\hline
$\delta$&$n$&$f=0.5$&$f=0.75$&$f=0.9$&$f=0.95$&$f=0.99$&$f=1$\\
\hline
\multirow{3}{1em}{$0$} &$250$&$0$&$0$&$0.005$&$0$&$0.005$&$0.01$\\
&$350$&$0.005$&$0$&$0.005$&$0$&$0.005$&$0.005$\\
&$500$&$0.015$&$0.005$&$0$&$0$&$0.005$&$0$\\
\hline
\multirow{3}{1em}{$0.1$} &$250$&$0.2$&$0.32$&$0.275$&$0.285$&$0.28$&$0.215$\\
&$350$&$0.385$&$0.445$&$0.54$&$0.45$&$0.445$&$0.43$\\
&$500$&$0.575$&$0.705$&$0.735$&$0.77$&$0.72$&$0.6$\\
\hline
\multirow{3}{1em}{$0.2$} &$250$&$0.845$&$0.975$&$0.955$&$0.94$&$0.955$&$0.895$\\
&$350$&$0.96$&$1$&$1$&$1$&$0.995$&$0.935$\\
&$500$&$0.985$&$1$&$1$&$1$&$1$&$0.955$\\
\hline
\multirow{3}{1em}{$0.3$} &$250$&$1$&$1$&$1$&$0.995$&$0.995$&$0.97$\\
&$350$&$1$&$1$&$1$&$1$&$1$&$0.97$\\
&$500$&$1$&$1$&$1$&$1$&$1$&$0.975$\\
\hline
\multirow{3}{1em}{$0.4$} &$250$&$1$&$1$&$1$&$1$&$0.985$&$0.965$\\
&$350$&$1$&$1$&$1$&$1$&$1$&$0.975$\\
&$500$&$1$&$1$&$1$&$1$&$1$&$0.98$\\
\hline
\multirow{3}{1em}{$0.5$} &$250$&$1$&$1$&$1$&$1$&$0.995$&$0.98$\\
&$350$&$1$&$1$&$1$&$1$&$1$&$0.99$\\
&$500$&$1$&$1$&$1$&$1$&$1$&$0.995$\\
\hline
\end{tabular}
\caption[Results for the highly correlated alternative group scenario]{\label{tab:sparse}Empirical rejection rate at level $5\%$ for the sparse alternative using single-carving.}
\end{table}
These results are to be compared to \citet[Table 3]{guo2019group} for $\delta$ in $\left\{0,0.2, 0.3\right\}$, where
they test the six methods in the sparse scenario. 
Their proposed methods $\phi_{\Sigma}\left(0.5\right)$ and 
$\phi_{\Sigma}\left(1\right)$ have lower power than our method for
$\delta=0.2$, while error control works very reliably for
all three methods. 
The power is (almost) at $1$ for all methods (except for their method $\phi_{I}$) 
for $\delta = 0.3$. 
The methods $\phi_{\text{hdi}}$ and $\phi_{\text{FD}}$ obtain values of power comparable to our method at a price of clearly higher error. 

\subsubsection{Multicarving for group testing: sparse scenario}

The results for multicarving are illustrated in Table \ref{tab:sparsemc}.
As for single-carving, error control in the highly correlated sparser alternative is no issue with the multicarve method. Namely, no ERR above $1.5\%$ occurs for $\delta=0$ for multicarving either. Again, using a selection fraction of $f=0.5$ seems to be favorable for multicarving.
\begin{table}[!htb]
\centering
\begin{tabular}{|lr|ccccc|}
\hline
$\delta$&$n$&$f=0.5$&$f=0.75$&$f=0.9$&$f=0.95$&$f=0.99$\\
\hline
\multirow{3}{1em}{$0$} &$250$&$0.005$&$0$&$0.005$&$0$&$0$\\
&$350$&$0.015$&$0$&$0.005$&$0$&$0$\\
&$500$&$0.005$&$0$&$0.005$&$0$&$0$\\
\hline
\multirow{3}{1em}{$0.1$} &$250$&$0.33$&$0.325$&$0.265$&$0.24$&$0.245$\\
&$350$&$0.57$&$0.495$&$0.42$&$0.38$&$0.37$\\
&$500$&$0.785$&$0.765$&$0.69$&$0.675$&$0.605$\\
\hline
\multirow{3}{1em}{$0.2$} &$250$&$0.99$&$0.985$&$0.98$&$0.945$&$0.955$\\
&$350$&$1$&$1$&$0.995$&$1$&$0.99$\\
&$500$&$1$&$1$&$1$&$1$&$1$\\
\hline
\multirow{3}{1em}{$0.3$} &$250$&$1$&$1$&$1$&$1$&$1$\\
&$350$&$1$&$1$&$1$&$1$&$1$\\
&$500$&$1$&$1$&$1$&$1$&$1$\\
\hline
\multirow{3}{1em}{$0.4$} &$250$&$1$&$1$&$1$&$1$&$1$\\
&$350$&$1$&$1$&$1$&$1$&$1$\\
&$500$&$1$&$1$&$1$&$1$&$1$\\
\hline
\multirow{3}{1em}{$0.5$} &$250$&$1$&$1$&$1$&$1$&$1$\\
&$350$&$1$&$1$&$1$&$1$&$1$\\
&$500$&$1$&$1$&$1$&$1$&$1$\\
\hline
\end{tabular}
\caption[Results for the highly correlated alternative group scenario]{\label{tab:sparsemc}Empirical rejection rate at level $5\%$ for the sparse alternative using multicarving.}
\end{table}

Looking at Table \ref{tab:sparse}, one sees that none of the single-carving configurations outperforms multicarving with $f=0.5$ in any scenario with $\delta>0$. Therefore, we can state that multicarving brings an improvement in this alternative as well when choosing the tuning parameters properly

\subsection{Effect of the aggregation parameter on the runtime} \label{app:quant}
Using a larger $\gamma_{min}$ for the aggregation in \eqref{eq:optquant} is 
favorable for computational reasons. First, only variables present in
at least $\gamma_{min} B$ models have to be tested for. The higher this
threshold is, the more variables can be omitted directly, reducing
computing time. Second, if we account for the multiplicity correction that
we impose through considering multiple variables and aggregating over
multiple splits, raw p-values of
$\dfrac{\alpha\gamma_{min}}{\tilde{s}\left(1-\text{log}\left(\gamma_{min}\right)\right)}$
or smaller should be possible. Otherwise, one can never observe a
significant effect occurring from
$P_j=\left(1-\text{log}\left(\gamma_{min}\right)\right)
Q_j\left(\gamma_{min}\right)$ (cf.\ Section \ref{ms}). Accordingly, we need
at least 
\begin{equation}\label{eq:minsample}
\dfrac{\tilde{s}\left(1-\text{log}\left(\gamma_{min}\right)\right)}{\alpha\gamma_{min}}
\end{equation}
MCMC samples to use the method to full capacity. This requirement decreases in $\gamma_{min}$ and is about $11$ times higher for $\gamma_{min}=0.05$ than for $\gamma_{min}=0.3$.

\subsection{Details for runtime considerations}\label{app:runtime}
We discuss what influences the runtime of multicarving and how to further speed it up.
Especially, we want to assess how the runtime behaves as $p \gg n \rightarrow \infty$.

We first review the structure of our method. For a total of $B$ times, the data is split into two parts, a model is selected on the first part, and p-values are calculated using the carving idea. For those p-values, a separate calculation for all of the $\tilde{s}$ selected variables is necessary. Lastly, the $B$ p-values of the different splits are aggregated per covariate. We ignore splitting the data, the initial selection stage, and the aggregation for our considerations since the computational bottleneck is the MCMC sampling required to calculate p-values.

Naturally, the runtime scales linearly in $B$. For every split, $\tilde{s}$ MCMC chains have to be sampled and one needs $\mathcal{O}\left( \tilde{s}/\alpha\right)$ samples in order to have the possibility to observe a significant result. Multicarving takes $B\tfrac{\left(1-\text{log}\left(\gamma_{min}\right)\right)}{\gamma_{min}}$ 
times as long as single-carving due to using multiple splits and the aggregation 
over the different splits (cf.\ Equation \eqref{eq:minsample}). Though, in 
practice convergence of the chain is another issue such that for single-carving 
more than the minimally required samples are likely to be generated and the 
difference between the two methods is slightly reduced. For single-carving and multicarving, there is a factor of $\tilde{s}^2$ involved as one needs $\tilde{s}$ chains of size $\mathcal{O}\left( \tilde{s}/\alpha\right)$. Lastly, sampling happens in a $\text{min}\left(\tilde{s}+1,n_{2}+1\right)$-dimensional space subject to $\tilde{s}$ inequality constraints (cf.\ Appendix \ref{app:mcmc}). 
We discuss two algorithms in the Appendix \ref{app:MCMC} and the choice of the 
MCMC algorithm influences the runtime. 

\cite{pakman2014exact} state that for their algorithm the exact run time also depends on the shape of the constraint such that a general statement cannot be made. There are steps of complexity $\mathcal{O}\big(\text{min}\left(\tilde{s}+1,n_{2}+1\right)^2\big)$ and $\mathcal{O}\big(\text{min}\left(\tilde{s}+1,n_{2}+1\right)\tilde{s}\big)$ involved, which can be bounded by $\mathcal{O}\left(\tilde{s}\right)^2$. However, the number of such calculations needed depends on the selection event's geometry.

For the hit-and-run algorithm adapted from the GitHub repository cited in \cite{fithian2014optimal}, every step involves solving a problem of the complexity of pure post-selection inference as in \cite{lee2016exact}. Due to the matrix equation involved in calculating the bounds, this leads to a complexity of $\mathcal{O}\left(\text{min}\left(\tilde{s}+1,n_{2}+1\right)\tilde{s}\right)\leq \mathcal{O}\left(\tilde{s}^2\right)$.

For both algorithms, we come up with an approximate bound of $\mathcal{O}\left(B\EE\left[{\tilde{s}^4}\right]\right)$ for multicarving where the expectation is due to the fact that $\tilde{s}$ is non-constant over splits.

In comparison, if we use the saturated viewpoint instead, p-values for every variable are determined by calculating bounds once taking at most $\mathcal{O}\left(n_1 \tilde{s}\right)$ steps. Assuming $\tilde{s}= \mathcal{O}\left(n_1\right)$, the inference process can be bounded by $\mathcal{O}\left(B\EE\left[{\tilde{s}^3}\right]\right)$ such that a factor of $\tilde{s}$ is saved. Though, it might be less appropriate to ignore the initial Lasso selection for runtime considerations in the saturated model.

Notably, there are several ways to speed up multicarving algorithmically. We want to state the two most obvious. As mentioned in Section \ref{app:quant}, not all covariates have to be tested for but only the ones selected in at least $\gamma_{min}B$ of the splits. This means that the algorithm described in Section \ref{mdc} has to be adjusted to selecting $B$ models first and performing inference afterwards, while the final outcome is not altered by this change. This improvement is more pronounced for higher values of $\gamma_{min}$. The exact same adjustment could also be applied to multisplitting. Second, not every MCMC chain has to be run to the full extent as in Equation \eqref{eq:minsample}. If it is already clear with fewer iterates that a covariate cannot be shown to be significant, the chain can be aborted in an earlier stage as for p-values clearly above the significance level the precision is less important.
\end{document}